\documentclass[11pt]{article}
\usepackage{graphicx}
\usepackage{amsmath}
\usepackage{amssymb}
\usepackage{caption2}
\begin{document}
\bibliographystyle{prsty}
\begin{center}
{\large {\bf \sc{  Revisit assignments of the new excited $\Omega_c$ states  with QCD sum rules }}} \\[2mm]
Zhi-Gang  Wang\footnote{E-mail: zgwang@aliyun.com. }, Xing-Ning Wei, Ze-Hui Yan    \\
 Department of Physics, North China Electric Power University,
Baoding 071003, P. R. China
\end{center}

\begin{abstract}
In this article, we distinguish  the contributions of the positive parity and negative parity  $\Omega_c$ states, study the masses and pole residues of the 1S, 1P, 2S and 2P
 $\Omega_c$ states with the spin $J=\frac{1}{2}$ and $\frac{3}{2}$ using the QCD sum rules in a consistent way, and revisit the assignments of the new narrow excited $\Omega_c^0$ states.
The predictions support assigning the $\Omega_c(3000)$
 to be the 1P $\Omega_c$ state with $J^P={\frac{1}{2}}^-$,
  assigning the $\Omega_c(3090)$ to be the 1P $\Omega_c$ state with $J^P={\frac{3}{2}}^-$ or the 2S $\Omega_c$ state with $J^P={\frac{1}{2}}^+$,   and
  assigning $\Omega_c(3119)$  to be the 2S $\Omega_c$ state with $J^P={\frac{3}{2}}^+$.
\end{abstract}



\section{Introduction}

Recently, the LHCb collaboration studied the  $\Xi_c^+ K^-$  mass spectrum and observed five new narrow excited $\Omega_c$ states,
$\Omega_c(3000)$, $\Omega_c(3050)$, $\Omega_c(3066)$, $\Omega_c(3090)$, $\Omega_c(3119)$ \cite{LHCb-Omega}. The measured masses and widths are
\begin{flalign}
 & \Omega_c(3000) : M = 3000.4 \pm 0.2 \pm 0.1 \mbox{ MeV}\, , \, \Gamma = 4.5\pm0.6\pm0.3 \mbox{ MeV} \, , \nonumber \\
 & \Omega_c(3050) : M = 3050.2 \pm 0.1 \pm 0.1 \mbox{ MeV}\, , \, \Gamma = 0.8\pm0.2\pm0.1 \mbox{ MeV} \, , \nonumber \\
 & \Omega_c(3066) : M = 3065.6 \pm 0.1 \pm 0.3 \mbox{ MeV}\, , \, \Gamma = 3.5\pm0.4\pm0.2 \mbox{ MeV} \, , \nonumber \\
 & \Omega_c(3090) : M = 3090.2 \pm 0.3 \pm 0.5 \mbox{ MeV}\, , \, \Gamma = 8.7\pm1.0\pm0.8 \mbox{ MeV} \, , \nonumber \\
 & \Omega_c(3119) : M = 3119.1 \pm 0.3 \pm 0.9 \mbox{ MeV}\, , \, \Gamma = 1.1\pm0.8\pm0.4 \mbox{ MeV} \, .
\end{flalign}
There have been several assignments for those new $\Omega_c$ states, such as the
2S $\Omega_c$  states  with $J^P={\frac{1}{2}}^+$ and ${\frac{3}{2}}^+$ \cite{Azizi-Omega,Cheng-Omega,LiuXiang-Omega,Azizi-Width-Omega}, the P-wave $\Omega_c$
 states with $J^P={\frac{1}{2}}^-$, ${\frac{3}{2}}^-$ or ${\frac{5}{2}}^-$ \cite{Cheng-Omega,LiuXiang-Omega,Azizi-Width-Omega,Chen-Omega,Rosner-Omega,WangZG-Omega,
WangZhu-Omega,Mathur-Omega,Zhong-Omega,Aliev-Omega}, the pentaquark states or molecular
pentaquark states with $J^P={\frac{1}{2}}^-$, ${\frac{3}{2}}^-$ or ${\frac{5}{2}}^-$ \cite{Ping-Omega,Polyakov-Omega,AnCS-Omega},
or the D-wave $\Omega_c$ states \cite{ZhangAL-Omega}.

In Refs.\cite{Azizi-Omega,Azizi-Width-Omega},   Agaev, Azizi and Sundu construct the interpolating currents without introducing the relative P-wave to
study the $\Omega_c$ states by taking into account the 1S, 1P, 2S states with $J=\frac{1}{2}$ and $\frac{3}{2}$ in the pole contributions in the QCD sum rules.
They use the 1S state plus continuum model to obtain the masses and pole residues of the 1S states firstly, then take them as input parameters and use the 1S state plus 1P
state plus continuum model to obtain the masses and pole residues of the 1P states, finally use the 1S state plus 1P state plus 2S state  plus continuum model to obtain the masses and pole
residues of the 2S states.
In Ref.\cite{Aliev-Omega},    Aliev, Bilmis and  Savci use the same interpolating currents to study the $\Omega_c$ states by taking into account the 1S and 1P states
with $J=\frac{1}{2}$ and $\frac{3}{2}$ in the pole contributions in the QCD sum rules. The potential quark models predict that the 1P and 2S $\Omega_c$ states have the
masses about $3.0-3.2\,\rm{GeV}$   \cite{Roberts2007,Ebert2011}. If the 1P and 2S $\Omega_c$ states lie in the same energy region, it is difficult to distinguish
their contributions in the QCD sum rules \cite{Azizi-Omega,Azizi-Width-Omega,Aliev-Omega}.

In Refs.\cite{WangPLB-Omega12,WangEPJC-Omega32,Wang-Omega-Negative,WangHbaryon}, we  construct the interpolating currents without introducing the relative P-wave
 to study the  $J^P={1\over 2}^{\pm}$ and ${3\over 2}^{\pm}$ heavy, doubly-heavy and triply-heavy baryon states    with the QCD sum rules  in a systematic way  by subtracting
 the contributions from the corresponding $J^P={1\over 2}^{\mp}$ and ${3\over 2}^{\mp}$  heavy, doubly-heavy and triply-heavy baryon states, and obtain satisfactory
 results.
In Ref.\cite{WangZG-Omega}, we   study the new excited $\Omega_c$ states  with the QCD sum rules  by introducing an explicit  P-wave involving  the two  $s$ quarks,
the predictions support assigning the $\Omega_c(3050)$, $\Omega_c(3066)$, $\Omega_c(3090)$ and $\Omega_c(3119)$  to be the  P-wave baryon states with $J^P={\frac{1}{2}}^-$,
 ${\frac{3}{2}}^-$, ${\frac{3}{2}}^-$ and ${\frac{5}{2}}^-$, respectively.

In this article, we distinguish the contributions of the S-wave (positive parity) and P-wave (negative parity) $\Omega_c$ states, study the masses and pole residues of the 1S, 1P, 2S and 2P
$\Omega_c$ states with the spin $J=\frac{1}{2}$ and $\frac{3}{2}$ using the QCD sum rules in details, and revisit the assignments of the new narrow excited $\Omega_c^0$ states.

 The article is arranged as follows:  we derive the QCD sum rules for the masses and pole residues of  the S-wave and P-wave ${\frac{1}{2}}$ and ${\frac{3}{2}}$ $\Omega_c$
 states in Sect.2;  in Sect.3, we present the numerical results and discussions; and Sect.4 is reserved for our conclusion.

\section{QCD sum rules for  the ${\frac{1}{2}}^\pm$ and ${\frac{3}{2}}^\pm$ $\Omega_c$ states}

Firstly, we write down  the two-point correlation functions  $\Pi(p)$ and  $\Pi_{\alpha\beta}(p)$  in the QCD sum rules,
\begin{eqnarray}
\Pi(p)&=&i\int d^4x e^{ip \cdot x} \langle0|T\left\{\eta(x)\bar{\eta}(0)\right\}|0\rangle \, , \nonumber\\
\Pi_{\alpha\beta}(p)&=&i\int d^4x e^{ip \cdot x} \langle0|T\left\{\eta_{\alpha}(x)\bar{\eta}_{\beta}(0)\right\}|0\rangle \, ,
\end{eqnarray}
where
\begin{eqnarray}
\eta(x)&=& \varepsilon^{ijk}  s^T_i(x)C\gamma_\alpha s_j(x) \gamma_5 \gamma^\alpha c_k(x)  \, ,  \nonumber \\
\eta_\alpha(x)&=& \varepsilon^{ijk}  s^T_i(x)C\gamma_\alpha s_j(x)   c_k(x)  \, ,
\end{eqnarray}
 the $i$, $j$ and $k$ are color indexes, and the $C$ is the charge conjugation
matrix.  In this article, we choose  the simple Ioffe type interpolating currents.

At the hadron side,  we  insert  a complete set  of intermediate $\Omega_c$  states with the
same quantum numbers as the current operators  $\eta(x)$, $i\gamma_5 \eta(x)$, $\eta_{\alpha}(x)$ and
$i\gamma_5 \eta_{\alpha}(x)$ into the correlation functions
$\Pi(p)$ and $\Pi_{\alpha\beta}(p)$ to obtain the hadronic representation
\cite{SVZ79,PRT85}. We isolate the pole terms of the lowest
1S, 1P, 2S and 2P $\Omega_c$ states ($\Omega_c$ and $\Omega_c^\prime$),   and obtain the  results:
\begin{eqnarray}
\Pi(p) & = & {\lambda^{+}_{\frac{1}{2}}}^2  {\!\not\!{p}+ M_{+} \over M_{+}^{2}-p^{2}  }
+  {\lambda^{-}_{\frac{1}{2}}}^2  {\!\not\!{p}- M_{-} \over M_{-}^{2}-p^{2}  } +{\lambda^{\prime+}_{\frac{1}{2}}}^2  {\!\not\!{p}+ M^\prime_{+} \over M_{+}^{\prime2}-p^{2}  }
+  {\lambda^{\prime-}_{\frac{1}{2}}}^2  {\!\not\!{p}- M_{-}^{\prime} \over M_{-}^{\prime2}-p^{2}  } +\cdots  \, , \nonumber\\
&=&\Pi_{\frac{1}{2}}(p^2)+\cdots\, ,
\end{eqnarray}
\begin{eqnarray}
\Pi_{\alpha\beta}(p) & = & \left({\lambda^{+}_{\frac{3}{2}}}^2  {\!\not\!{p}+ M_{+} \over M_{+}^{2}-p^{2}  }
+  {\lambda^{-}_{\frac{3}{2}}}^2  {\!\not\!{p}- M_{-} \over M_{-}^{2}-p^{2}  } +{\lambda^{\prime+}_{\frac{3}{2}}}^2  {\!\not\!{p}+ M^\prime_{+} \over M_{+}^{\prime2}-p^{2}  }
+  {\lambda^{\prime-}_{\frac{3}{2}}}^2  {\!\not\!{p}- M_{-}^{\prime} \over M_{-}^{\prime2}-p^{2}  } \right) \nonumber\\
&& \left(- g_{\alpha\beta}+\frac{\gamma_\alpha\gamma_\beta}{3}+\frac{2p_\alpha p_\beta}{3p^2}-\frac{p_\alpha\gamma_\beta-p_\beta \gamma_\alpha}{3\sqrt{p^2}}
\right) +\cdots  \, ,\nonumber\\
&=&\Pi_{\frac{3}{2}}(p^2)\,\left(- g_{\alpha\beta}\right)+\cdots\, .
\end{eqnarray}

The currents  $\eta(0)$ and $\eta_{\alpha}(0)$ couple potentially to the spin-parity   $J^P={\frac{1}{2}}^\pm$ and  ${\frac{3}{2}}^\pm$  $\Omega_c$
 states   $\Omega_{\frac{1}{2}}^{(\prime)\pm}$ and   $\Omega_{\frac{3}{2}}^{(\prime)\pm}$, respectively \cite{WangHbaryon,Wang-2625-2815,Oka96,WangPc},
\begin{eqnarray}
\langle 0| \eta (0)|\Omega_{\frac{1}{2}}^{(\prime)+}(p)\rangle &=&\lambda^{(\prime)+}_{\frac{1}{2}} U^{+}(p,s) \, ,  \nonumber\\
\langle 0| \eta_\alpha (0)|\Omega_{\frac{3}{2}}^{(\prime)+}(p)\rangle &=&\lambda^{(\prime)+}_{\frac{3}{2}} U^{+}_{\alpha }(p,s) \, ,\\
\langle 0| \eta (0)|\Omega_{\frac{1}{2}}^{(\prime)-}(p)\rangle &=&\lambda^{(\prime)-}_{\frac{1}{2}} i\gamma_5 U^{-}(p,s) \, ,  \nonumber\\
\langle 0| \eta_\alpha (0)|\Omega_{\frac{3}{2}}^{(\prime)-}(p)\rangle &=&\lambda^{(\prime)-}_{\frac{3}{2}} i\gamma_5 U^{-}_{\alpha }(p,s) \, ,
\end{eqnarray}
where the $\lambda^{(\prime)\pm}_{\frac{1}{2}}$ and $\lambda^{(\prime)\pm}_{\frac{3}{2}}$ are the pole residues or the current-baryon couplings,
the spinors $U^{\pm}(p,s)$ and $U^{\pm}_{\alpha}(p,s)$ satisfy the relations,
\begin{eqnarray}
\sum_s U(p,s) \overline{U}(p,s)&=& \!\not\!{p}+M_{\pm}^{(\prime)} \,  , \nonumber \\
\sum_s U_\alpha(p,s) \overline{U}_\beta(p,s)&=&\left(\!\not\!{p}+M_{\pm}^{(\prime)}\right)\left( -g_{\alpha\beta}+\frac{\gamma_\alpha\gamma_\beta}{3}+\frac{2p_\alpha p_\beta}{3p^2}-\frac{p_\alpha\gamma_\beta-p_\beta \gamma_\alpha}{3\sqrt{p^2}} \right) \,  ,
\end{eqnarray}
and $p^2=M^{(\prime)2}_{\pm}$ on  mass-shell, the $s$ are the polarizations or spin indexes of the spinors, and should  be distinguished from the $s$ quark or the energy $s$.

We obtain the hadronic spectral densities at hadron  side through dispersion relation,
\begin{eqnarray}
\frac{{\rm Im}\Pi_{j}(s)}{\pi}&=&\!\not\!{p} \left[{\lambda^{+}_{j}}^2 \delta\left(s-M_{+}^2\right)+{\lambda^{-}_{j}}^2 \delta\left(s-M_{-}^2\right)
+{\lambda^{\prime+}_{j}}^2 \delta\left(s-M_{+}^{\prime2}\right)+{\lambda^{\prime-}_{j}}^2 \delta\left(s-M_{-}^{\prime2}\right)\right] \nonumber\\
&&+\left[M_{+}{\lambda^{+}_{j}}^2 \delta\left(s-M_{+}^2\right)-M_{-}{\lambda^{-}_{j}}^2 \delta\left(s-M_{-}^2\right)+M_{+}^{\prime}{\lambda^{\prime+}_{j}}^2 \delta\left(s-M_{+}^{\prime2}\right)\right.\nonumber\\
&&\left.-M_{-}^{\prime}{\lambda^{\prime-}_{j}}^2 \delta\left(s-M_{-}^{\prime2}\right)\right]  \nonumber\\
&=&\!\not\!{p}\, \rho^1_{j,H}(s)+\rho^0_{j,H}(s) \, ,
\end{eqnarray}
where $j=\frac{1}{2}$, $\frac{3}{2}$, the subscript $H$ denotes  the hadron side,
then we introduce the weight function $\exp\left(-\frac{s}{T^2}\right)$  to obtain the QCD sum rules at the hadron side,
\begin{eqnarray}
\int_{m_c^2}^{s_0}ds \left[\sqrt{s}\rho^1_{j,H}(s)+\rho^0_{j,H}(s)\right]\exp\left( -\frac{s}{T^2}\right)
&=&2M_{+}{\lambda^{+}_{j}}^2\exp\left( -\frac{M_{+}^2}{T^2}\right)\nonumber\\
&&+2M_{+}^\prime{\lambda^{\prime+}_{j}}^2\exp\left( -\frac{M_{+}^{\prime2}}{T^2}\right) \, ,  \\
\int_{m_c^2}^{s_0}ds \left[\sqrt{s}\rho^1_{j,H}(s)-\rho^0_{j,H}(s)\right]\exp\left( -\frac{s}{T^2}\right)
&=&2M_{-}{\lambda^{-}_{j}}^2\exp\left( -\frac{M_{-}^2}{T^2}\right)\nonumber\\
&&+2M_{-}^\prime{\lambda^{\prime-}_{j}}^2\exp\left( -\frac{M_{-}^{\prime2}}{T^2}\right) \, ,
\end{eqnarray}
where the $s_0$ are the continuum thresholds and the $T^2$ are the Borel parameters \cite{WangPc}. We distinguish
the contributions of the positive parity and negative  parity $\Omega_c$ states    unambiguously according to Eqs.(9-11).

At the QCD side, we  calculate the light quark parts of the correlation functions
 $\Pi(p)$ and $\Pi_{\alpha\beta}(p)$ with the full light quark propagators $S_{ij}(x)$  in the coordinate space \cite{Pascual-1984},
 \begin{eqnarray}
S_{ij}(x)&=& \frac{i\delta_{ij}\!\not\!{x}}{ 2\pi^2x^4}-\frac{\delta_{ij}m_s}{4\pi^2x^2}-\frac{\delta_{ij}\langle
\bar{s}s\rangle}{12} +\frac{i\delta_{ij}\!\not\!{x}m_s\langle\bar{s}s\rangle}{48}-\frac{\delta_{ij}x^2\langle \bar{s}g_s\sigma Gs\rangle}{192}\nonumber\\
&&+\frac{i\delta_{ij}x^2\!\not\!{x} m_s\langle \bar{s}g_s\sigma Gs\rangle }{1152} -\frac{ig_s G^{a}_{\alpha\beta}t^a_{ij}(\!\not\!{x}
\sigma^{\alpha\beta}+\sigma^{\alpha\beta} \!\not\!{x})}{32\pi^2x^2} -\frac{1}{8}\langle\bar{s}_j\sigma^{\mu\nu}s_i \rangle \sigma_{\mu\nu} +\cdots \, , \nonumber \\
\end{eqnarray}
 and take   the full $c$-quark propagator $C_{ij}(x)$ in the momentum space \cite{PRT85},
\begin{eqnarray}
C_{ij}(x)&=&\frac{i}{(2\pi)^4}\int d^4k e^{-ik \cdot x} \left\{
\frac{\delta_{ij}}{\!\not\!{k}-m_c}
-\frac{g_sG^n_{\alpha\beta}t^n_{ij}}{4}\frac{\sigma^{\alpha\beta}(\!\not\!{k}+m_c)+(\!\not\!{k}+m_c)
\sigma^{\alpha\beta}}{(k^2-m_c^2)^2}\right.\nonumber\\
&&\left. -\frac{g_s^2 (t^at^b)_{ij} G^a_{\alpha\beta}G^b_{\mu\nu}(f^{\alpha\beta\mu\nu}+f^{\alpha\mu\beta\nu}+f^{\alpha\mu\nu\beta}) }{4(k^2-m_c^2)^5}+\cdots\right\} \, , \end{eqnarray}
\begin{eqnarray}
f^{\alpha\beta\mu\nu}&=&(\!\not\!{k}+m_c)\gamma^\alpha(\!\not\!{k}+m_c)\gamma^\beta(\!\not\!{k}+m_c)\gamma^\mu(\!\not\!{k}+m_c)\gamma^\nu(\!\not\!{k}+m_c)\, ,
\end{eqnarray}
 $q=u,d,s$,  $t^n=\frac{\lambda^n}{2}$, the $\lambda^n$ is the Gell-Mann matrix. In Eq.(12), we add the term
  $\langle\bar{s}_j\sigma_{\mu\nu}s_i \rangle$  originates  from the Fierz re-ordering   of the
   $\langle s_i \bar{s}_j\rangle$ to  absorb the gluons  emitted from other quark lines to form
    $\langle\bar{s}_j g_s G^a_{\alpha\beta} t^a_{mn}\sigma_{\mu\nu} s_i \rangle$    to extract the mixed condensate     $\langle\bar{s}g_s\sigma G s\rangle$.
    The term $-\frac{1}{8}\langle\bar{s}_j\sigma^{\mu\nu}s_i \rangle \sigma_{\mu\nu}$ was introduced in Ref.\cite{WangTetraquark}.
We compute  the integrals both in the coordinate space and momentum space  to obtain the correlation functions $\Pi_{j}(p^2)$, then obtain  the QCD spectral densities
 through  dispersion relation,
\begin{eqnarray}
\frac{{\rm Im}\Pi_{j}(s)}{\pi}&=&\!\not\!{p}\, \rho^1_{j,QCD}(s)+\rho^0_{j,QCD}(s) \, ,
\end{eqnarray}
where $j=\frac{1}{2}$, $\frac{3}{2}$, the explicit expressions of the QCD spectral densities $\rho^1_{j,QCD}(s)$ and  $\rho^0_{j,QCD}(s)$
 can be rewritten in a concise form after multiplying the weight function $\exp\left(-\frac{s}{T^2}\right)$ to obtain the integrals
$ \int_{m_c^2}^{\infty}ds\, \sqrt{s}\,\rho^1_{j,QCD}(s)\exp\left(-\frac{s}{T^2}\right)$ and $\int_{m_c^2}^{\infty}ds\, \rho^0_{j,QCD}(s)\exp\left(-\frac{s}{T^2}\right)$.

We  take the quark-hadron duality,  introduce the continuum thresholds  $s_0$ and the weight function $\exp\left(-\frac{s}{T^2}\right)$  to obtain  the  QCD sum rules:
\begin{eqnarray}
\int_{m_c^2}^{s_0}ds \left[\sqrt{s}\rho^1_{j,H}(s)+\rho^0_{j,H}(s)\right]\exp\left( -\frac{s}{T^2}\right)
&=& \int_{m_c^2}^{s_0}ds \left[\sqrt{s}\rho^1_{j,QCD}(s)+\rho^0_{j,QCD}(s)\right]\nonumber\\
&&\exp\left( -\frac{s}{T^2}\right)\, ,  \\
\int_{m_c^2}^{s_0}ds \left[\sqrt{s}\rho^1_{j,H}(s)-\rho^0_{j,H}(s)\right]\exp\left( -\frac{s}{T^2}\right)
&=& \int_{m_c^2}^{s_0}ds \left[\sqrt{s}\rho^1_{j,QCD}(s)-\rho^0_{j,QCD}(s)\right]\nonumber\\
&&\exp\left( -\frac{s}{T^2}\right)\,  ,
\end{eqnarray}
where  $j=\frac{1}{2}$, $\frac{3}{2}$,
\begin{eqnarray}
\rho^0_{j,QCD}(s)&=&m_c\,\rho^{0}_{j}(s)\, ,\nonumber\\
\rho^1_{j,QCD}(s)&=&\rho^{1}_{j}(s)\, ,
\end{eqnarray}
\begin{eqnarray}
\rho^{0}_{\frac{1}{2}}(s)&=&\frac{3}{32\pi^4}\int_{x_i}^1dx \, (1-x)^2(s-\widetilde{m}_c^2)^2 -\frac{3m_s\langle\bar{s}s\rangle}{2\pi^2}\int_{x_i}^1dx   \nonumber\\
&&+\frac{1}{32\pi^2}\langle\frac{\alpha_sGG}{\pi}\rangle\int_{x_i}^1dx \frac{(1-x)^2}{x^2}\left[1 -\frac{s}{3} \delta(s-\widetilde{m}_c^2)\right] \nonumber\\
&&+\frac{1}{32\pi^2}\langle\frac{\alpha_sGG}{\pi}\rangle\int_{x_i}^1dx \,\frac{2-3x}{x} \nonumber\\
&&-\frac{m_s\langle\bar{s}g_s\sigma Gs\rangle}{4\pi^2}\int_{x_i}^1dx \,\frac{1}{x}\,\delta(s-\widetilde{m}_c^2)  \nonumber\\
&&+\frac{5m_s\langle\bar{s}g_s\sigma Gs\rangle}{12\pi^2}\delta(s-m_c^2)+\frac{4\langle\bar{s}s\rangle^2}{3}\delta(s-m_c^2)
\nonumber\\
&&+\frac{\langle\bar{s} s\rangle \langle\bar{s}g_s\sigma Gs\rangle}{3T^2}\left( 1-\frac{2m_c^2}{T^2}\right) \,\delta(s-m_c^2) \nonumber\\
&&-\frac{ \langle\bar{s}g_s\sigma Gs\rangle^2}{12T^4} \left( 1+\frac{3m_c^2}{T^2}-\frac{m_c^4}{T^4}\right)\,\delta(s-m_c^2)  \, ,
\end{eqnarray}

\begin{eqnarray}
\rho^{1}_{\frac{1}{2}}(s)&=&\frac{1}{16\pi^4}\int_{x_i}^1dx \, x(1-x)^3(5s-3\widetilde{m}_c^2)(s-\widetilde{m}_c^2)  -\frac{m_s\langle\bar{s}s\rangle}{\pi^2}\int_{x_i}^1dx x  \nonumber\\
&&+\frac{m_s\langle\bar{s}s\rangle}{\pi^2}\int_{x_i}^1dx x(1-x)\left[3+s\,\delta\,(s-\widetilde{m}_c^2)  \right]  \nonumber\\
&&+\frac{1}{48\pi^2}\langle\frac{\alpha_sGG}{\pi}\rangle\int_{x_i}^1dx \left\{(1-x)\left[3+s\,\delta\,(s-\widetilde{m}_c^2)  \right]+(1-2x) \right\} \nonumber\\
&&-\frac{m_c^2}{72\pi^2}\langle\frac{\alpha_sGG}{\pi}\rangle\int_{x_i}^1dx \frac{(1-x)^3}{x^2}\left(1 +\frac{s}{2T^2} \right)\delta(s-\widetilde{m}_c^2)  \nonumber\\
&&-\frac{m_s\langle\bar{s}g_s\sigma Gs\rangle}{3\pi^2}\int_{x_i}^1dx x \left(1+\frac{s}{2T^2}  \right) \delta\,(s-\widetilde{m}_c^2) \nonumber\\
&&-\frac{m_s\langle\bar{s}g_s\sigma Gs\rangle}{8\pi^2}\int_{x_i}^1dx \, \delta(s-\widetilde{m}_c^2)\nonumber\\
&&+\frac{m_s\langle\bar{s}g_s\sigma Gs\rangle}{4\pi^2}\delta(s-m_c^2)+\frac{2\langle\bar{s}s\rangle^2}{3}\delta(s-m_c^2)\nonumber\\
&&-\frac{\langle\bar{s}s\rangle \langle\bar{s}g_s\sigma Gs\rangle }{6T^2}\left(1+\frac{2m_c^2}{T^2} \right)\delta(s-m_c^2)\nonumber\\
&&-\frac{m_c^2 \langle\bar{s}g_s\sigma Gs\rangle^2}{24T^6} \left(1-\frac{m_c^2}{T^2} \right) \,\delta(s-m_c^2) \, ,
\end{eqnarray}

\begin{eqnarray}
\rho^{0}_{\frac{3}{2}}(s)&=&\frac{1}{64\pi^4}\int_{x_i}^1dx \, (x+2)(1-x)^2(s-\widetilde{m}_c^2)^2 -\frac{m_s\langle\bar{s}s\rangle}{4\pi^2}\int_{x_i}^1dx(2-x)   \nonumber\\
&&-\frac{m_c^2}{576\pi^2}\langle\frac{\alpha_sGG}{\pi}\rangle\int_{x_i}^1dx \frac{(x+2)(1-x)^2}{x^3}\delta(s-\widetilde{m}_c^2)\nonumber\\
&& +\frac{1}{192\pi^2}\langle\frac{\alpha_sGG}{\pi}\rangle\int_{x_i}^1dx \left[\frac{(x+2)(1-x)^2}{x^2}-(2-x)\right]  \nonumber\\
&&+\frac{m_s\langle\bar{s}g_s\sigma Gs\rangle}{24\pi^2}\int_{x_i}^1dx\delta(s-\widetilde{m}_c^2)+\frac{m_s\langle\bar{s}g_s\sigma Gs\rangle}{12\pi^2} \,\delta(s-m_c^2)  \nonumber\\
&&+\frac{\langle\bar{s}s\rangle^2}{3}\delta(s-m_c^2) -\frac{m_c^2\langle\bar{s}s\rangle \langle\bar{s}g_s\sigma Gs\rangle }{6T^4}\delta(s-m_c^2)\nonumber\\
&& -\frac{m_c^2  \langle\bar{s}g_s\sigma Gs\rangle^2 }{24T^6}\left(1-\frac{m_c^2}{2T^2} \right)\delta(s-m_c^2) \, ,
\end{eqnarray}

\begin{eqnarray}
\rho^{1}_{\frac{3}{2}}(s)&=&\frac{1}{64\pi^4}\int_{x_i}^1dx \, x(x+2)(1-x)^2(s-\widetilde{m}_c^2)^2 -\frac{m_s\langle\bar{s}s\rangle}{4\pi^2}\int_{x_i}^1dx x(2-x)   \nonumber\\
&&-\frac{m_c^2}{576\pi^2}\langle\frac{\alpha_sGG}{\pi}\rangle\int_{x_i}^1dx \frac{(x+2)(1-x)^2}{x^2}\delta(s-\widetilde{m}_c^2)\nonumber\\
&&-\frac{1}{192\pi^2}\langle\frac{\alpha_sGG}{\pi}\rangle\int_{x_i}^1dx \,x(2-x)  \nonumber\\
&&+\frac{m_s\langle\bar{s}g_s\sigma Gs\rangle}{24\pi^2}\int_{x_i}^1dx x \delta(s-\widetilde{m}_c^2)+\frac{m_s\langle\bar{s}g_s\sigma Gs\rangle}{12\pi^2} \,\delta(s-m_c^2)  \nonumber\\
&&+\frac{\langle\bar{s}s\rangle^2}{3}\delta(s-m_c^2) -\frac{\langle\bar{s}s\rangle \langle\bar{s}g_s\sigma Gs\rangle }{6T^2}\left(1+\frac{m_c^2}{T^2} \right)\delta(s-m_c^2)\nonumber\\
&& +\frac{m_c^4  \langle\bar{s}g_s\sigma Gs\rangle^2 }{48T^8} \delta(s-m_c^2)\, ,
\end{eqnarray}
$\widetilde{m}_c^2=\frac{m_c^2}{x}$, $x_i=\frac{m_c^2}{s}$.

The QCD sum rules can be written more explicitly,
\begin{eqnarray}
2M_{+}{\lambda^{+}_{j}}^2\exp\left( -\frac{M_{+}^2}{T^2}\right)+2M_{+}^{\prime}{\lambda^{\prime+}_{j}}^2\exp\left( -\frac{M_{+}^{\prime2}}{T^2}\right)
&=& \int_{m_c^2}^{s_0}ds \left[\sqrt{s}\rho^1_{j,QCD}(s)+\rho^0_{j,QCD}(s)\right]\nonumber\\
&&\exp\left( -\frac{s}{T^2}\right)\, ,  \\
2M_{-}{\lambda^{-}_{j}}^2\exp\left( -\frac{M_{-}^2}{T^2}\right)+2M_{-}^{\prime}{\lambda^{\prime-}_{j}}^2\exp\left( -\frac{M_{-}^{\prime2}}{T^2}\right)
&=& \int_{m_c^2}^{s_0}ds \left[\sqrt{s}\rho^1_{j,QCD}(s)-\rho^0_{j,QCD}(s)\right]\nonumber\\
&&\exp\left( -\frac{s}{T^2}\right)\, .
\end{eqnarray}
The contributions of the positive parity and negative  parity $\Omega_c$ states are separated explicitly.

Firstly, we choose low continuum threshold parameters $s_0$ so as not to include the contributions of the 2S and 2P $\Omega_c$ states ($\Omega_c^\prime$),
 and  obtain the QCD sum rules for
 the masses of the 1S and 1P $\Omega_c$ states,
 \begin{eqnarray}
 M^2_{+} &=& \frac{-\frac{d}{d(1/T^2)}\int_{m_c^2}^{s_0}ds \left[\sqrt{s}\rho^1_{j,QCD}(s)+\rho^0_{j,QCD}(s)\right]\exp\left( -\frac{s}{T^2}\right)}{\int_{m_c^2}^{s_0}ds \left[\sqrt{s}\rho^1_{j,QCD}(s)+\rho^0_{j,QCD}(s)\right]\exp\left( -\frac{s}{T^2}\right)}\, , \\
  M^2_{-} &=& \frac{-\frac{d}{d(1/T^2)}\int_{m_c^2}^{s_0}ds \left[\sqrt{s}\rho^1_{j,QCD}(s)-\rho^0_{j,QCD}(s)\right]\exp\left( -\frac{s}{T^2}\right)}{\int_{m_c^2}^{s_0}ds \left[\sqrt{s}\rho^1_{j,QCD}(s)-\rho^0_{j,QCD}(s)\right]\exp\left( -\frac{s}{T^2}\right)}\, ,
\end{eqnarray}
then obtain  the pole residues $\lambda^{+}_{j}$ and $\lambda^{-}_{j}$.

Now we take the masses and pole residues of the 1S and 1P $\Omega_c$ states as input parameters,
and postpone the continuum threshold parameters $s_0$ to larger values to include the contributions of the 2S and 2P $\Omega_c$ states,
and obtain the QCD sum rules for
 the masses of the 2S and 2P $\Omega_c$ states,
\begin{eqnarray}
 M^{\prime2}_{+} &=& \frac{-\frac{d}{d(1/T^2)}\left\{\int_{m_c^2}^{s_0}ds \left[\sqrt{s}\rho^1_{j,QCD}(s)+\rho^0_{j,QCD}(s)\right]
 \exp\left( -\frac{s}{T^2}\right)-2M_{+}{\lambda^{+}_{j}}^2
 \exp\left( -\frac{M_{+}^2}{T^2}\right)\right\}}
 {\int_{m_c^2}^{s_0}ds \left[\sqrt{s}\rho^1_{j,QCD}(s)+\rho^0_{j,QCD}(s)\right]\exp\left( -\frac{s}{T^2}\right)-2M_{+}{\lambda^{+}_{j}}^2
 \exp\left( -\frac{M_{+}^2}{T^2}\right)}\, , \nonumber\\
 \end{eqnarray}
 \begin{eqnarray}
 M^{\prime2}_{-} &=& \frac{-\frac{d}{d(1/T^2)}\left\{\int_{m_c^2}^{s_0}ds \left[\sqrt{s}\rho^1_{j,QCD}(s)-\rho^0_{j,QCD}(s)\right]
 \exp\left( -\frac{s}{T^2}\right)-2M_{-}{\lambda^{-}_{j}}^2\exp\left( -\frac{M_{-}^2}{T^2}\right)\right\}}{\int_{m_c^2}^{s_0}ds
 \left[\sqrt{s}\rho^1_{j,QCD}(s)-\rho^0_{j,QCD}(s)\right]\exp\left( -\frac{s}{T^2}\right)-2M_{-}{\lambda^{-}_{j}}^2\exp\left( -\frac{M_{-}^2}{T^2}\right)}\, , \nonumber\\
\end{eqnarray}
then obtain  the pole residues $\lambda^{\prime+}_{j}$ and $\lambda^{\prime-}_{j}$.

\section{Numerical results and discussions}
The input parameters   are taken to be the standard values
$\langle\bar{q}q \rangle=-(0.24\pm 0.01\, \rm{GeV})^3$,  $\langle\bar{s}s \rangle=(0.8\pm0.1)\langle\bar{q}q \rangle$,
   $\langle\bar{s}g_s\sigma G s \rangle=m_0^2\langle \bar{s}s \rangle$,
$m_0^2=(0.8 \pm 0.1)\,\rm{GeV}^2$, $\langle \frac{\alpha_s
GG}{\pi}\rangle=(0.33\,\rm{GeV})^4 $    at the energy scale  $\mu=1\, \rm{GeV}$
\cite{SVZ79,PRT85,Ioffe-PPNP,ColangeloReview}, $m_{c}(m_c)=(1.275\pm0.025)\,\rm{GeV}$ and $m_s(\mu=2\,\rm{GeV})=(0.095\pm0.005)\,\rm{GeV}$
 from the Particle Data Group \cite{PDG}.
 The updated values from the Particle Data Group in version 2016 \cite{PDG} are slightly different from the corresponding ones in version 2014, we take the old values to
   make consistent predictions with the same parameters and criteria chosen in previous works.  If we choose the updated values
   $m_{c}(m_c)=(1.28\pm0.03)\,\rm{GeV}$ and $m_s(\mu=2\,\rm{GeV})=0.096^{+0.008}_{-0.004}\,\rm{GeV}$ \cite{PDG}, the central value of the predicted mass of the $\Omega_c({\rm 1S})$
   is  $2.6991\,\rm{GeV}$ rather than $2.6983\,\rm{GeV}$, the predicted mass presented  in Table 2 survives, so the old values are OK.
   The values of the $m_0^2$, $\langle\bar{s}s \rangle/\langle\bar{q}q \rangle$ and $\langle\bar{s}g_s\sigma Gs \rangle/\langle\bar{q}g_s \sigma Gq \rangle$ vary
    in rather large ranges from different theoretical determinations, for example, in Ref.\cite{Aladashvili},  $\langle\bar{s}g_s\sigma Gs \rangle/\langle\bar{q}g_s \sigma Gq \rangle=0.95 \pm 0.15$, which differs from
    the standard value $\langle\bar{s}g_s\sigma Gs \rangle/\langle\bar{q}g_s \sigma Gq \rangle=\langle\bar{s}s \rangle/\langle\bar{q}q \rangle=0.8 \pm 0.1$ remarkably \cite{Ioffe-PPNP}. In this article,
    we take the standard values or the old values still accepted now \cite{Ioffe-PPNP,ColangeloReview}.

 We take into account
the energy-scale dependence of  the input parameters from the renormalization group equation,
\begin{eqnarray}
\langle\bar{s}s \rangle(\mu)&=&\langle\bar{s}s \rangle(Q)\left[\frac{\alpha_{s}(Q)}{\alpha_{s}(\mu)}\right]^{\frac{4}{9}}\, , \nonumber\\
\langle\bar{s}g_s \sigma Gs \rangle(\mu)&=&\langle\bar{s}g_s \sigma Gs \rangle(Q)\left[\frac{\alpha_{s}(Q)}{\alpha_{s}(\mu)}\right]^{\frac{2}{27}}\, , \nonumber\\
m_s(\mu)&=&m_s({\rm 2GeV} )\left[\frac{\alpha_{s}(\mu)}{\alpha_{s}({\rm 2GeV})}\right]^{\frac{4}{9}} \, ,\nonumber\\
m_c(\mu)&=&m_c(m_c)\left[\frac{\alpha_{s}(\mu)}{\alpha_{s}(m_c)}\right]^{\frac{12}{25}} \, ,\nonumber\\
\alpha_s(\mu)&=&\frac{1}{b_0t}\left[1-\frac{b_1}{b_0^2}\frac{\log t}{t} +\frac{b_1^2(\log^2{t}-\log{t}-1)+b_0b_2}{b_0^4t^2}\right]\, ,
\end{eqnarray}
  where $t=\log \frac{\mu^2}{\Lambda^2}$, $b_0=\frac{33-2n_f}{12\pi}$, $b_1=\frac{153-19n_f}{24\pi^2}$, $b_2=\frac{2857-\frac{5033}{9}n_f+\frac{325}{27}n_f^2}{128\pi^3}$,
   $\Lambda=213\,\rm{MeV}$, $296\,\rm{MeV}$  and  $339\,\rm{MeV}$ for the flavors  $n_f=5$, $4$ and $3$, respectively  \cite{PDG},
   and evolve all the input parameters to the optimal energy scales  $\mu$ to extract the masses of the $\Omega_c$ states.
          The energy scale dependence of the  quark masses and quark condensates
   is known beyond the leading order, the energy scale dependence of the  mixed quark condensates is only known in the leading order \cite{Narison-mix,Narison-Book}.
      In this article, we take the leading order approximation
   in a consistent way, and take the energy scale dependence of the mixed condensates  presented in Refs.\cite{Narison-mix,Narison-Book}, while a quite 
   different energy scale
   dependence of the mixed condensates is presented in Refs.\cite{Aladashvili,Beneke-Dosch}. It is interesting to take the energy scale dependence presented in Refs.\cite{Aladashvili,Beneke-Dosch}, this 
   may be our next work.  
    For the heavy degrees of freedom, we take the favors $n_f=4$,
   the power in the $m_c(\mu)$ is $\frac{12}{25}$. For the light degrees of freedom, we take the flavors $n_f=3$, the powers in the
   $\langle\bar{s}s \rangle(\mu)$, $\langle\bar{s}g_s \sigma Gs \rangle(\mu)$ and
   $m_s(\mu)$ are $\frac{4}{9}$ (or $\frac{12}{27}$), $\frac{2}{27}$ and $\frac{4}{9}$,  respectively. If we take the favors $n_f=4$, the powers in the
   $\langle\bar{s}s \rangle(\mu)$, $\langle\bar{s}g_s \sigma Gs \rangle(\mu)$ and
   $m_s(\mu)$ are   $\frac{12}{25}$ , $\frac{2}{25}$ and $\frac{12}{25}$,  respectively, in fact, the induced tiny difference in numerical calculations can be neglected.
   As far as the fine constant $\alpha_s(\mu)$ is concerned, we choose the next-to-next-to-leading order approximation, which is consistent with the values determined
   experimentally \cite{PDG}.

In Fig.1, we plot the correlation functions  $\Pi_{j,+}$ and $\Pi_{j,-}$ with variations of the energy scales $\mu$ and the Borel parameters $T^2$,
\begin{eqnarray}
\Pi_{j,+}&=& \int_{m_c^2}^{\infty}ds \left[\sqrt{s}\rho^1_{j,QCD}(s)+\rho^0_{j,QCD}(s)\right]\exp\left( -\frac{s}{T^2}\right)\, ,  \\
\Pi_{j,-}&=& \int_{m_c^2}^{\infty}ds \left[\sqrt{s}\rho^1_{j,QCD}(s)-\rho^0_{j,QCD}(s)\right]\exp\left( -\frac{s}{T^2}\right)\, .
\end{eqnarray}
From the figure, we can see that the $\Pi_{j,+}$ and $\Pi_{j,-}$ increase remarkably with increase of the energy scale $\mu$ at the region $T^2>4.0\,\rm{GeV}^2$,
 while at the region $T^2<3.0\,\rm{GeV}^2$, the $\Pi_{j,+}$ and $\Pi_{j,-}$ increase slowly with increase of the energy scale  $\mu$. All in all,  we cannot obtain energy scale independent QCD sum rules,
some constraints are needed to determine the energy scales of the QCD spectral densities in a consistent way.

 Now we take a short digression to discuss how to choose the optimal energy scales.  In the heavy quark limit, the heavy quark $Q$ serves as a static well potential and  combines with a  light quark $q$  to form a heavy diquark  in  color antitriplet,
 or combines with a light diquark in color antitriplet to form a heavy baryon in color singlet.
 The heavy antiquark $\overline{Q}$  serves  as another static well potential and combines with a  light antiquark $\bar{q}^\prime$  to form a heavy antidiquark  in  color
 triplet,  or combines with a light antidiquark  in color triplet to form a heavy antibaryon in color singlet. Then the heavy diquark and heavy antidiquark combine together to form a hidden-charm or hidden-bottom
 tetraquark state.
The heavy baryons  $B$ and tetraquark states $X/Y/Z$  are characterized by the effective heavy quark masses
${\mathbb{M}}_Q$ (or constituent quark masses) and the virtuality  $V=\sqrt{M^2_{B}-{\mathbb{M}}_Q^2}$, $\sqrt{M^2_{X/Y/Z}-(2{\mathbb{M}}_Q)^2}$
 (or bound energy not as robust). The diquark-quark type baryon states  and
  diquark-antidiquark type tetraquark states  are expected to have the same
  effective $Q$-quark masses ${\mathbb{M}}_Q$, which embody  the net effects of the complex dynamics \cite{WangTetraquark,Wang-Xi3080}.
  In Refs.\cite{WangTetraquark,WangMolecule}, we study the acceptable energy scales of the QCD spectral densities  for
 the hidden-charm (hidden-bottom) tetraquark states and molecular states   in the QCD sum rules in details  for the first time,
 and suggest an energy scale formula  $\mu=\sqrt{M^2_{X/Y/Z}-(2{\mathbb{M}}_Q)^2}$ by setting $\mu=V$ to determine  the optimal   energy scales with
  the effective heavy quark masses ${\mathbb{M}}_Q$.

We fit  the   effective $c$-quark mass ${\mathbb{M}}_{c}$ to reproduce the experimental value of the mass of the $Z^\pm_c(3900)$  in the
scenario of  tetraquark  state \cite{WangTetraquark}.
In this article, we use the  empirical energy scale formula
$ \mu =\sqrt{M_{\Omega_c}^2-{\mathbb{M}}_c^2}$ to determine
the optimal energy scales of the QCD spectral densities, and take the updated value of the effective $c$-quark mass ${\mathbb{M}}_c=1.82\,\rm{GeV}$ \cite{WangEPJC4260}.
For detailed discussions about the energy scale formula $ \mu =\sqrt{M_{\Omega_c}^2-{\mathbb{M}}_c^2}$, one can consult Ref.\cite{Wang-Xi3080}. According to the energy scale formula
$ \mu =\sqrt{M_{\Omega_c}^2-{\mathbb{M}}_c^2}$, we extract the masses of the ground states (see Eqs.(25-26)) and the first radial excited states (see Eqs.(27-28)) at different
energy scales.

In Fig.2, we plot the masses  and pole residues of the $\Omega_c({\rm 1S},\frac{1}{2})$,  $\Omega_c({\rm 1S},\frac{3}{2})$,  $\Omega_c({\rm 1P},\frac{1}{2})$ and
      $\Omega_c({\rm 1P},\frac{3}{2})$ with variations of the energy scale $\mu$\,  for the central values of the Borel parameters  and threshold parameters
       shown in Table 1. From the figure, we can see that the predicted masses  decrease   monotonously but mildly with increase of the energy scale $\mu$,   the constraint
       $ \mu =\sqrt{M_{\Omega_c}^2-{\mathbb{M}}_c^2}$ is not difficult  to satisfy. On the other hand, the pole residues increase monotonously and mildly with increase
       of the energy scale $\mu$, which is consistent with Fig.1, as the Borel parameters are chosen as $T^2<3.0\,\rm{GeV}^2$. At the vicinities of the energy scales presented in Table 1,
       the uncertainties induced by the uncertainties of the energy scales are tiny.

For the $Z_c(3900)$, the uncertainty of the energy scale of the QCD spectral density is about $\delta\mu=0.1\,\rm{GeV}$, the uncertainty of the
  effective $c$-quark mass ${\mathbb{M}}_c$  can be estimated as $\delta{\mathbb{M}}_c =\frac{\mu_0}{4{\mathbb{M}}_c}\delta \mu=0.02\,\rm{GeV}$ from the  equation,
\begin{eqnarray}
\mu&=&\sqrt{M_{X/Y/Z}^2-4\left({\mathbb{M}}_c\pm\delta{\mathbb{M}}_c\right)^2}=\sqrt{M_{X/Y/Z}^2-4{\mathbb{M}}_c^2}
\sqrt{1\mp \frac{8{\mathbb{M}}_c\delta{\mathbb{M}}_c}{M_{X/Y/Z}^2-4{\mathbb{M}}_c^2}}\nonumber\\
&=&\mu_0\left(1\mp \frac{4{\mathbb{M}}_c\delta{\mathbb{M}}_c}{\mu_0^2}\right)=\mu_0\mp\delta\mu\, ,
\end{eqnarray}
where the $\mu_0$ is the central value. The uncertainties $\delta\mu$ in this article can be estimated as
$\delta\mu=\frac{{\mathbb{M}}_c}{\mu_0}\delta{\mathbb{M}}_c<0.02\,\rm{GeV}$
from the equation,
\begin{eqnarray}
\mu&=&\sqrt{M_{\Omega_c}^2-\left({\mathbb{M}}_c\pm\delta{\mathbb{M}}_c\right)^2}=\mu_0\mp\frac{{\mathbb{M}}_c}{\mu_0}\delta{\mathbb{M}}_c\, .
\end{eqnarray}
 The predicted masses and pole residues are not sensitive to variations of the energy scales,
the small uncertainty $\delta{\mathbb{M}}_c= 0.02\,\rm{GeV}$ or $\delta\mu<0.02\,\rm{GeV}$ can be neglected safely.

\begin{figure}
 \centering
 \includegraphics[totalheight=5cm,width=7cm]{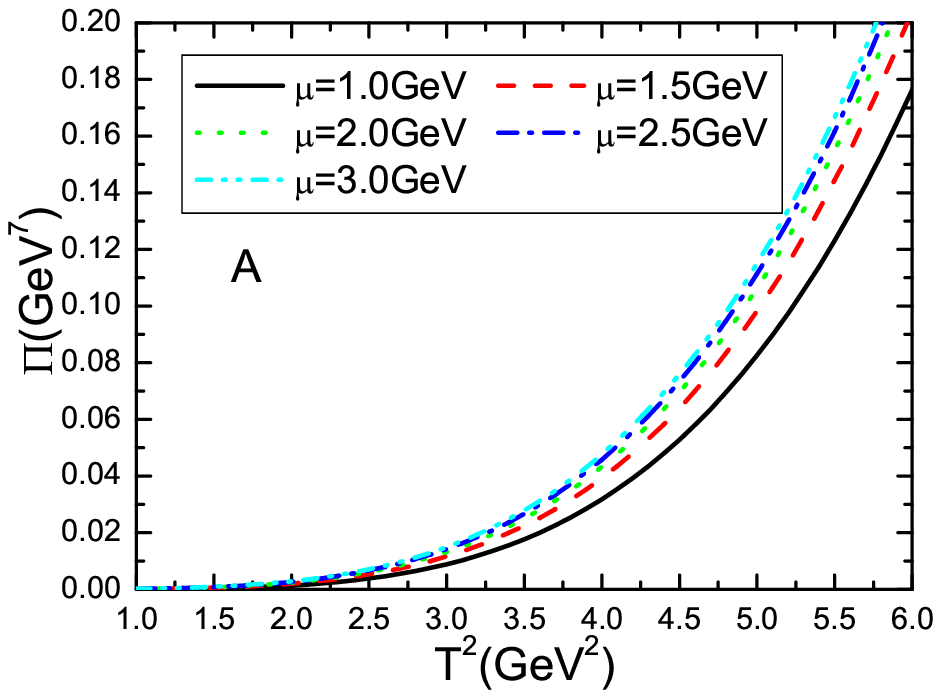}
 \includegraphics[totalheight=5cm,width=7cm]{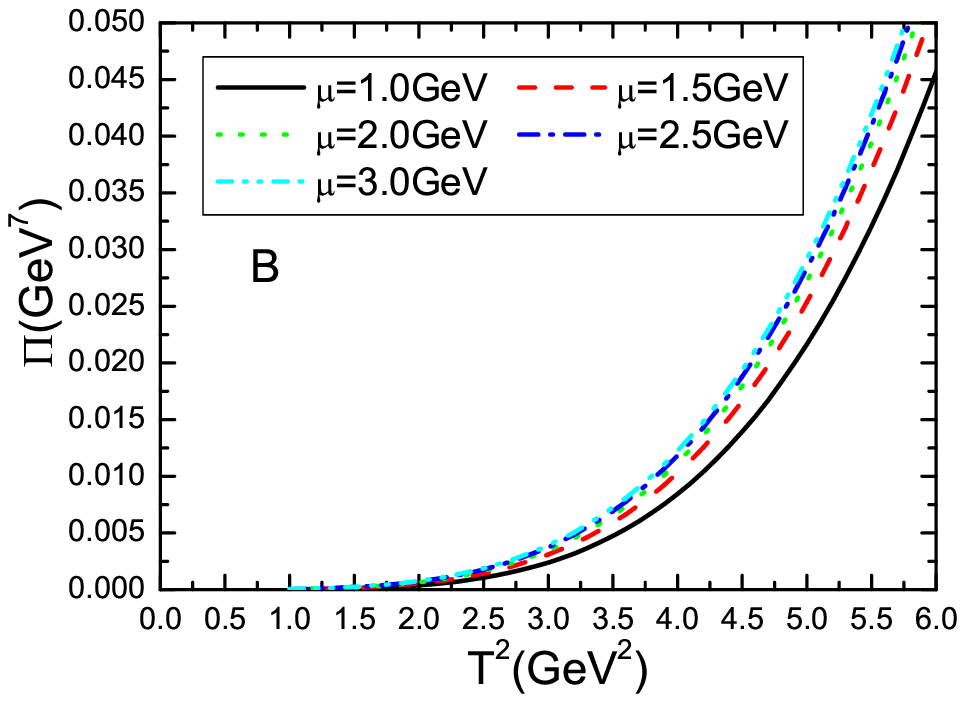}
 \includegraphics[totalheight=5cm,width=7cm]{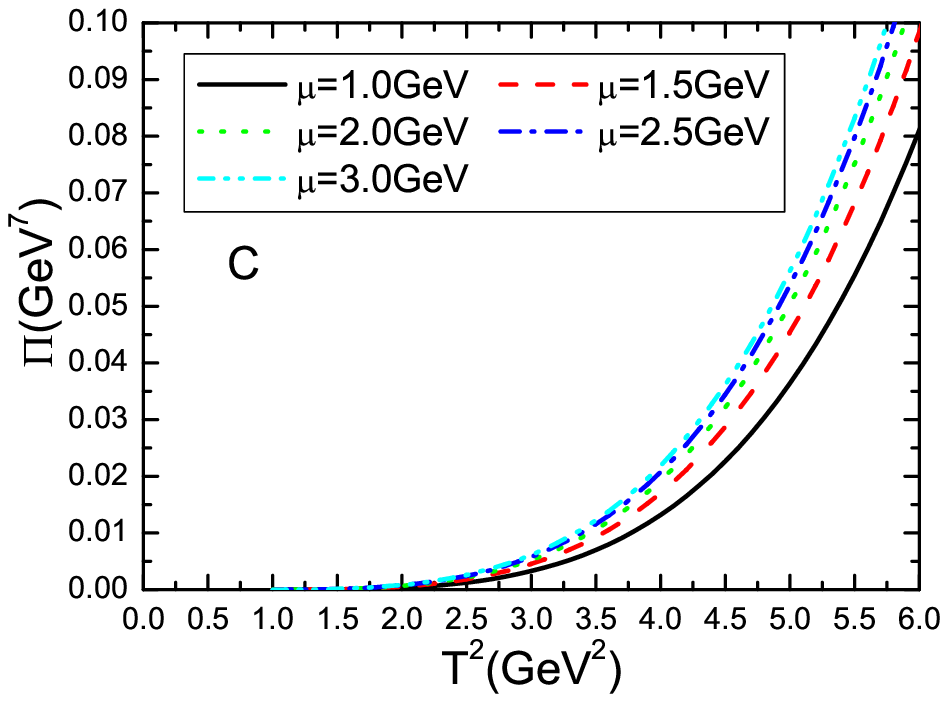}
 \includegraphics[totalheight=5cm,width=7cm]{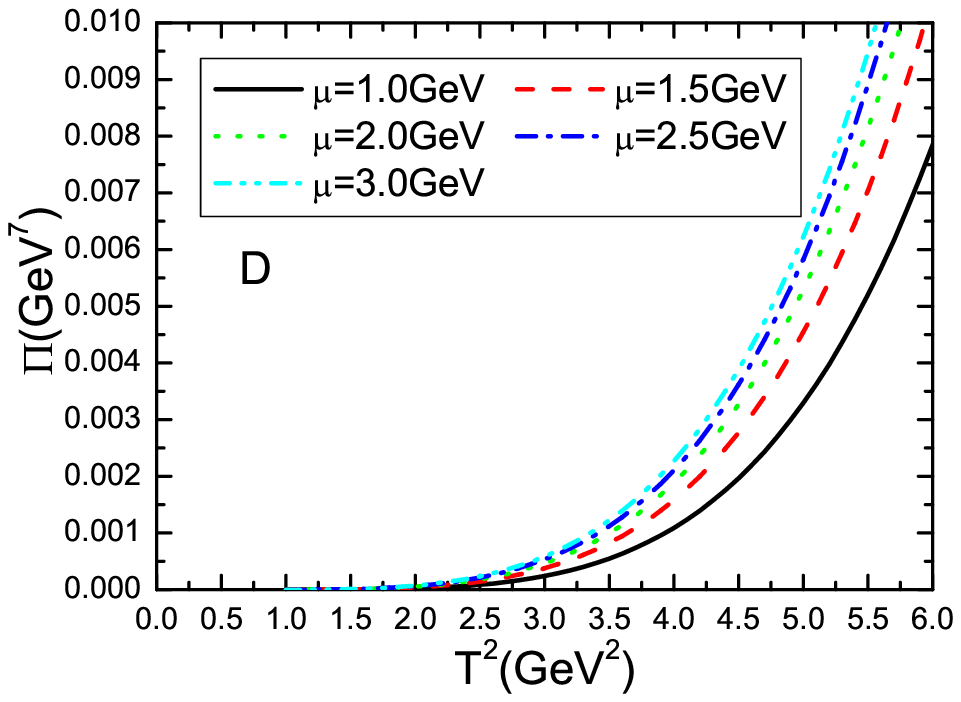}
         \caption{ The correlation functions   with variations of the energy scales $\mu$ and Borel parameters $T^2$,
         where the $A$, $B$, $C$ and $D$ correspond to the
         $\Pi_{\frac{1}{2},+}$, $\Pi_{\frac{3}{2},+}$, $\Pi_{\frac{1}{2},-}$ and $\Pi_{\frac{3}{2},-}$, respectively.  }
\end{figure}

We search for the ideal Borel parameters $T^2$ and continuum threshold parameters $s_0$   according to  the  four criteria:

$\bf{1_\cdot}$ Pole dominance at the hadron side, the pole contributions are about $(40-70)\%$;

$\bf{2_\cdot}$ Convergence of the operator product expansion, the dominant contributions come from the perturbative terms;

$\bf{3_\cdot}$ Appearance of the Borel platforms;

$\bf{4_\cdot}$ Satisfying the energy scale formula $ \mu =\sqrt{M_{\Omega_c}^2-{\mathbb{M}}_c^2}$,  \\
by try and error,  and present the optimal energy scales $\mu$,  ideal Borel parameters $T^2$, continuum threshold parameters $s_0$,
pole contributions and perturbative contributions in Table 1. From Table 1, we can see that the criteria $\bf{1}$ and $\bf{2}$ can be satisfied, the two basic criteria
of the QCD sum rules can be satisfied, and we expect to make reliable predictions.

We take into account all uncertainties  of the input    parameters,
and obtain  the masses and pole residues of
 the  1S, 1P, 2S and 2P $\Omega_c$ states, which are shown explicitly in
Table 2. From Table 2, we can see that the criterion $\bf{4}$ can be satisfied.  In Figs.3-4,
we plot the masses and pole residues of the 1S, 1P, 2S and 2P $\Omega_c$ states
 with variations
of the Borel parameters $T^2$ at much larger intervals   than the  Borel windows shown in Table 1. In the Borel windows, the uncertainties originate
from the Borel parameters $T^2$ are very small, the Borel platforms exist, the criterion $\bf{3}$ can be satisfied. Now the  four criteria are all satisfied, and we expect to make
reliable predictions. In the Borel windows, the uncertainties of the predicted masses are about $(3-5)\%$, as we obtain the masses from a ratio, see Eqs.(25-28), the
uncertainties
originate from a special parameter in the  numerator and denominator    cancel   out   with each other, so the net uncertainties are very small. On the other hand, the uncertainties
of the pole residues are about $(10-16)\%$, which are much larger. The uncertainties $\delta \lambda_{\Omega_c}$ are compatible with the uncertainties of the decay constants
$f_\pi=127\pm 15\,\rm{MeV}$ and $f_{\rho}=213\pm20\,\rm{MeV}$ from the QCD sum rules \cite{ColangeloReview}.

In  Table 2, we also present the experimental  values \cite{LHCb-Omega,PDG}.  The present predictions support assigning the $\Omega_c(3000)$
 to be the 1P $\Omega_c$ state with $J^P={\frac{1}{2}}^-$, assigning the $\Omega_c(3090)$ to be the 1P $\Omega_c$ state with $J^P={\frac{3}{2}}^-$ or the 2S $\Omega_c$
 state with $J^P={\frac{1}{2}}^+$,   and assigning the $\Omega_c(3119)$  to be the 2S $\Omega_c$ state with $J^P={\frac{3}{2}}^+$ (or the 1P $\Omega_c$ state with
 $J^P={\frac{5}{2}}^-$ \cite{WangZG-Omega}).
 The present predictions indicate that the
1P $\Omega_c$ state with $J^P={\frac{3}{2}}^-$ and the 2S $\Omega_c$ state with $J^P={\frac{1}{2}}^+$ have degenerate masses,
it is difficult to distinguish them by the masses  alone, we have to study their  strong decays.  Other predictions   can be confronted to the experimental data in the future.

\begin{figure}
 \centering
\includegraphics[totalheight=5cm,width=7cm]{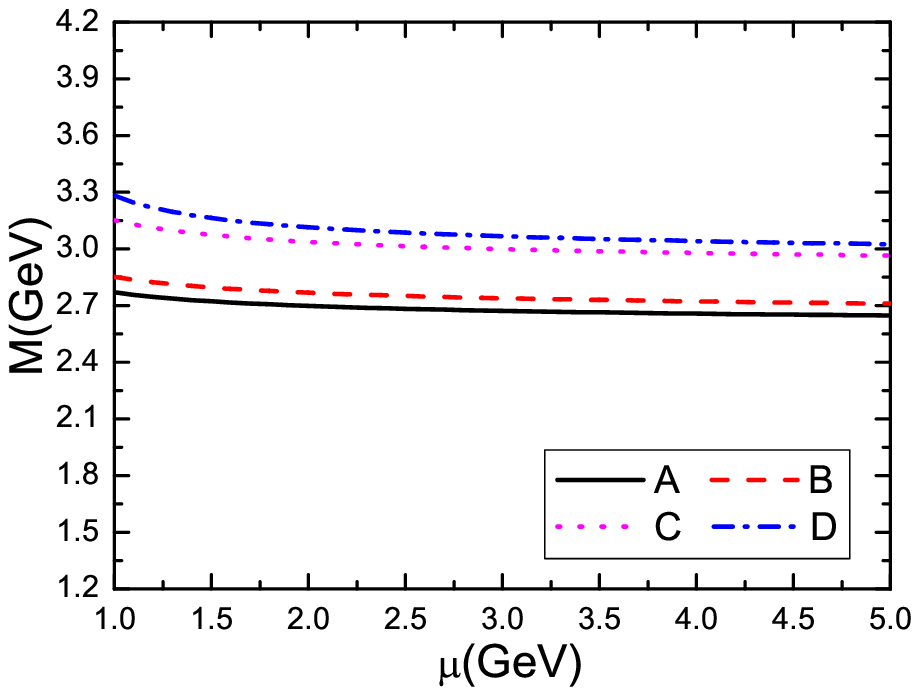}
\includegraphics[totalheight=5cm,width=7cm]{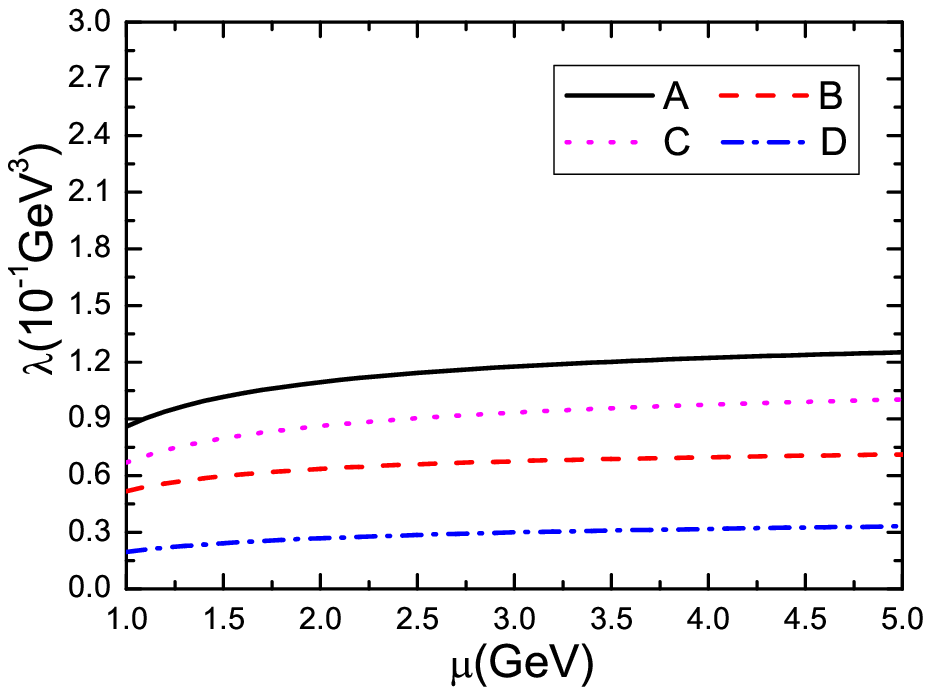}
       \caption{ The masses and pole residues of the $\Omega_c$ states  with variations of the energy scale $\mu$ for the central values of the Borel parameters  and threshold parameters shown in Table 1, where the $A$, $B$, $C$ and $D$ correspond to the $\Omega_c({\rm 1S},\frac{1}{2})$,  $\Omega_c({\rm 1S},\frac{3}{2})$,  $\Omega_c({\rm 1P},\frac{1}{2})$ and
         $\Omega_c({\rm 1P},\frac{3}{2})$, respectively.  }
\end{figure}

\begin{table}
\begin{center}
\begin{tabular}{|c|c|c|c|c|c|c|c|}\hline\hline
                      &$J^P$                &$\mu(\rm GeV)$  &$T^2 (\rm{GeV}^2)$  & $\sqrt{s_0} (\rm{GeV})$   & pole         & perturbative   \\  \hline
$\Omega_c({\rm 1S})$  &${\frac{1}{2}}^+$    &2.0            &$2.3-2.9$           & $3.30\pm0.10$             & $(41-69)\%$  & $(86-90)\%$  \\ \hline
$\Omega_c({\rm 1S})$  &${\frac{3}{2}}^+$    &2.1            &$2.4-3.0$           & $3.40\pm0.10$             & $(46-72)\%$  & $(87-91)\%$  \\ \hline

$\Omega_c({\rm 1P})$  &${\frac{1}{2}}^-$    &2.4            &$2.2-2.8$           & $3.40\pm0.10$             & $(40-68)\%$  & $(117-130)\%$  \\ \hline
$\Omega_c({\rm 1P})$  &${\frac{3}{2}}^-$    &2.5            &$2.2-2.8$           & $3.50\pm0.10$             & $(39-67)\%$  & $(106-114)\%$  \\ \hline

$\Omega_c({\rm 2S})$  &${\frac{1}{2}}^+$    &2.5            &$2.6-3.2$           & $3.45\pm0.10$             & $(43-68)\%$  & $(90-93)\%$  \\ \hline
$\Omega_c({\rm 2S})$  &${\frac{3}{2}}^+$    &2.5            &$2.7-3.3$           & $3.50\pm0.10$             & $(45-69)\%$  & $(90-93)\%$  \\ \hline

$\Omega_c({\rm 2P})$  &${\frac{1}{2}}^-$    &2.9            &$2.4-3.0$           & $3.70\pm0.10$             & $(53-78)\%$  & $(111-118)\%$  \\ \hline
$\Omega_c({\rm 2P})$  &${\frac{3}{2}}^-$    &2.9            &$2.4-3.0$           & $3.75\pm0.10$             & $(49-75)\%$  & $(104-108)\%$  \\ \hline\hline
\end{tabular}
\end{center}
\caption{ The optimal energy scales $\mu$, Borel parameters $T^2$, continuum threshold parameters $s_0$,
 pole contributions (pole)   and  perturbative contributions (perturbative) for the $\Omega_c$ states.}
\end{table}

\begin{table}
\begin{center}
\begin{tabular}{|c|c|c|c|c|c|c|c|}\hline\hline
                       &$J^P$                 &$M(\rm{GeV})$           &$\lambda (\rm{GeV}^3)$                       &(expt) (MeV)      \\ \hline
$\Omega_c({\rm 1S})$   &${\frac{1}{2}}^+$     &$2.70^{+0.11}_{-0.13}$  &$1.09^{+0.17}_{-0.15}\times 10^{-1}$         &2695.2    \\  \hline
$\Omega_c({\rm 1S})$   &${\frac{3}{2}}^+$     &$2.76^{+0.11}_{-0.12}$  &$0.64^{+0.09}_{-0.08}\times 10^{-1}$         &2765.9    \\  \hline

 $\Omega_c({\rm 1P})$  &${\frac{1}{2}}^-$     &$3.02^{+0.12}_{-0.07}$  &$0.90^{+0.13}_{-0.10}\times 10^{-1}$         &?\, 3000.4   \\  \hline
 $\Omega_c({\rm 1P})$  &${\frac{3}{2}}^-$     &$3.09^{+0.08}_{-0.06}$  &$0.29^{+0.04}_{-0.04}\times 10^{-1}$         &?\, 3090.2   \\  \hline

$\Omega_c({\rm 2S})$   &${\frac{1}{2}}^+$     &$3.09^{+0.11}_{-0.12}$  &$0.82^{+0.09}_{-0.09}\times 10^{-1}$         &?\, 3090.2   \\  \hline
$\Omega_c({\rm 2S})$   &${\frac{3}{2}}^+$     &$3.12^{+0.12}_{-0.12}$  &$0.37^{+0.03}_{-0.04}\times 10^{-1}$         &?\, 3119.1   \\  \hline

$\Omega_c({\rm 2P})$   &${\frac{1}{2}}^-$     &$3.40^{+0.10}_{-0.10}$  &$0.91^{+0.09}_{-0.09}\times 10^{-1}$         &  \\  \hline
$\Omega_c({\rm 2P})$   &${\frac{3}{2}}^-$     &$3.46^{+0.10}_{-0.11}$  &$0.27^{+0.04}_{-0.03}\times 10^{-1}$         &  \\  \hline

$\Omega_c({\rm 1P})$   &${\frac{5}{2}}^-$     &$3.11^{+0.10}_{-0.10}$  &$1.07^{+0.17}_{-0.17}\times 10^{-1}\rm{GeV}$ &?\, 3119.1  \\  \hline\hline
\end{tabular}
\end{center}
\caption{ The masses and pole residues of the $\Omega_c$ states, the masses are compared with the experimental data, the values of the $\Omega_c({\rm 1P})$  with
$J^P={\frac{5}{2}}^-$ are taken from Ref.\cite{WangZG-Omega}. }
\end{table}

In Refs.\cite{Azizi-Omega,Azizi-Width-Omega},   Agaev, Azizi and Sundu  study the $\Omega_c$ states by taking into account the 1S, 1P, 2S states with $J=\frac{1}{2}$ and $\frac{3}{2}$
in the pole contributions, and assign the $\Omega_c(3000)$, $\Omega_c(3050)$, $\Omega_c(3066)$ and $\Omega_c(3119)$ to
be the $({\rm 1P},{\frac{1}{2}}^-)$, $({\rm 1P}, {\frac{3}{2}}^-)$, $({\rm 2S},{\frac{1}{2}}^+)$ and $({\rm 2S}, {\frac{3}{2}}^+)$ states, respectively.
In Ref.\cite{Aliev-Omega},    Aliev, Bilmis and  Savci use the same interpolating currents to study the $\Omega_c$ states by taking into account
the 1S and 1P states with $J=\frac{1}{2}$ and $\frac{3}{2}$ in the pole contributions, and assign the $\Omega_c(3000)$ and $\Omega_c(3066)$   to
be the $({\rm 1P},{\frac{1}{2}}^-)$ and $({\rm 1P}, {\frac{3}{2}}^-)$ states, respectively.  In Refs.\cite{Azizi-Omega,Azizi-Width-Omega,Aliev-Omega}, the contributions of the
$\Omega_c$ states with  positive parity and negative parity are not separated explicitly, there are some contaminations from the 2S or 1P states.
In Ref.\cite{WangZG-Omega}, we separate the contributions of the positive parity and negative parity $\Omega_c$ states explicitly, and
   study the new excited $\Omega_c$ states  with the QCD sum rules  by introducing an explicit  P-wave involving   the two  $s$ quarks.
   The predictions support assigning the $\Omega_c(3050)$, $\Omega_c(3066)$, $\Omega_c(3090)$ and $\Omega_c(3119)$  to be the  P-wave $\Omega_c$ states
    with $J^P={\frac{1}{2}}^-$, ${\frac{3}{2}}^-$, ${\frac{3}{2}}^-$ and ${\frac{5}{2}}^-$, respectively.
    Compared with Refs.\cite{Azizi-Omega,Azizi-Width-Omega,Aliev-Omega}, the methods used in the present work and Ref.\cite{WangZG-Omega} have the advantage that the contributions
    of the  $\Omega_c$ states with  positive parity and negative parity are  separated explicitly, there are no contaminations from the 2S or 1P states.

In the diquark-quark models for the heavy baryon states, the angular momentum between the two light quarks is denoted by $L_\rho$,
while the angular momentum between the  light diquark and the heavy quark is denoted by $L_\lambda$.
In Refs.\cite{Azizi-Omega,Azizi-Width-Omega,Aliev-Omega} and present work, the currents with $L_\rho=L_\lambda=0$ are chosen to explore the P-wave $\Omega_c$ states,
although the currents couple  potentially  to the P-wave $\Omega_c$ states, we are unable to know the substructures of the P-wave $\Omega_c$ states, and cannot  distinguish
whether they have $L_\lambda=1$ or $L_\rho=1$. In Ref.\cite{WangZG-Omega}, we choose the currents with $L_\lambda=1$ to interpolate the $\Omega_c$ states, and obtain the
predicted masses $(3.06\pm0.11)\,\rm{GeV}$ and $(3.06\pm0.10)\,\rm{GeV}$ for the $J^P={\frac{3}{2}}^-$ $\Omega_c$ states with slightly different substructures,
which support assigning the  $\Omega_c(3066)$ and $\Omega_c(3090)$  to be the  P-wave $\Omega_c$ states
    with   $J^P={\frac{3}{2}}^-$ and $L_\lambda=1$. While in the present work, we obtain the mass $3.09^{+0.08}_{-0.06}\,\rm{GeV}$ for the $J^P={\frac{3}{2}}^-$ $\Omega_c$ state.
     If we take the central values of the predicted masses as  references,
    the $\Omega_c(3066)$ and  $\Omega_c(3090)$  can be tentatively assigned to
    be the ${\frac{3}{2}}^-$ $\Omega_c$ states with $L_\lambda=1$ and $L_\rho=1$, respectively. However, the assignment $\Omega_c(3090)=\rm (2S,{\frac{1}{2}}^+)$ is
   also possible  according
    to the predicted mass $3.09^{+0.11}_{-0.12}\,\rm{GeV}$ for the $\rm (2S,{\frac{1}{2}}^+)$ state.

Now we  summarize  the assignments based on the QCD sum rules in Table 3. From Table 3, we can see that all the calculations based on the QCD sum rules support assigning the $\Omega_c(3000)$ to be the
1P ${\frac{1}{2}}^-$ state, while the assignments of the other $\Omega_c$ states are under debate.  We have to study the decay widths to make the assignments on more
solid foundation.
In Ref.\cite{Azizi-Width-Omega},   Agaev, Azizi and Sundu  study the decays of the $\Omega_c$ states to the $\Xi_c^+K^-$ by calculating the hadronic coupling constants
$g_{\Omega_c\Xi_cK}$ with the light-cone  QCD sum rules, however, they
use an over simplified hadronic representation and neglect the contributions of the excited $\Xi_c$ states.

Experimentally, we can search  for those new excited $\Omega_c$ states through  strong decays and electromagnetic decays to the final states
$ \Xi_c^+K^-$, $\Xi_c^0 \bar{K}^0$, $\Xi_c^{\prime+}K^-$, $\Xi_c^{\prime0} \bar{K}^0$, $\Xi_c^{*+}K^-$, $\Xi_c^{*0} \bar{K}^0$, $\Xi^-D^+$,
$\Xi^0D^0$, $\Omega_c(2695) \gamma$, $\Omega_c(2770) \gamma$, and measure  the branching fractions precisely, which can shed light on the nature of  those $\Omega_c$ states.
More theoretical works on the partial decay widths based on the QCD sum rules are still needed.

\begin{table}
\begin{center}
\begin{tabular}{|c|c|c|c|c|c|c|c|}\hline\hline
                      &$J^P$                 &$nL$       & References                     \\ \hline
$\Omega_c(3000)$      &${\frac{1}{2}}^-$     &1P         & \cite{Azizi-Omega,Azizi-Width-Omega,WangZG-Omega,Aliev-Omega,Wang-Omega-Negative}\, and\, This\,\,Work   \\  \hline

$\Omega_c(3050)$      &${\frac{3}{2}}^-$     &1P         & \cite{Azizi-Omega,Azizi-Width-Omega}  \\  \hline

$\Omega_c(3050)$      &${\frac{1}{2}}^-$     &1P         & \cite{WangZG-Omega}   \\  \hline

$\Omega_c(3066)$      &${\frac{3}{2}}^-$     &1P         & \cite{WangZG-Omega,Aliev-Omega}   \\  \hline

$\Omega_c(3066)$      &${\frac{1}{2}}^+$     &2S         & \cite{Azizi-Omega,Azizi-Width-Omega}   \\  \hline

$\Omega_c(3090)$      &${\frac{3}{2}}^-$     &1P         & \cite{WangZG-Omega}\, and\, This\,\,Work   \\  \hline

$\Omega_c(3090)$      &${\frac{1}{2}}^+$     &2S         &   This\,\,Work   \\  \hline

$\Omega_c(3119)$      &${\frac{3}{2}}^+$     &2S         &   \cite{Azizi-Omega,Azizi-Width-Omega}\, and\, This\,\,Work   \\  \hline

$\Omega_c(3119)$      &${\frac{5}{2}}^-$     &1P         & \cite{WangZG-Omega}    \\  \hline\hline
\end{tabular}
\end{center}
\caption{ The possible assignments of the new $\Omega_c$ states based on the QCD sum rules. }
\end{table}

\begin{figure}
 \centering
 \includegraphics[totalheight=5cm,width=7cm]{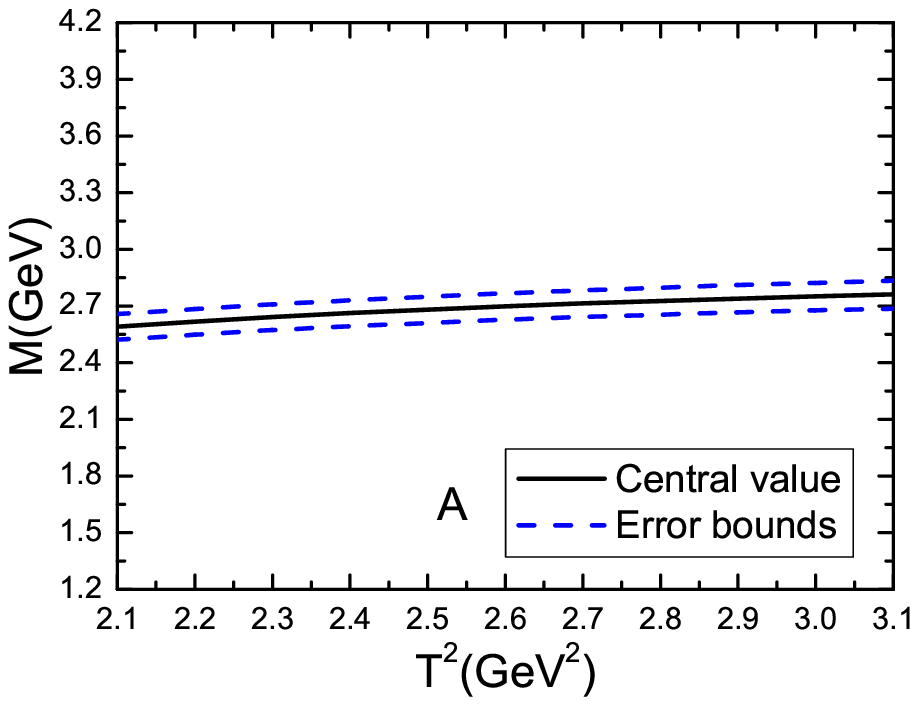}
 \includegraphics[totalheight=5cm,width=7cm]{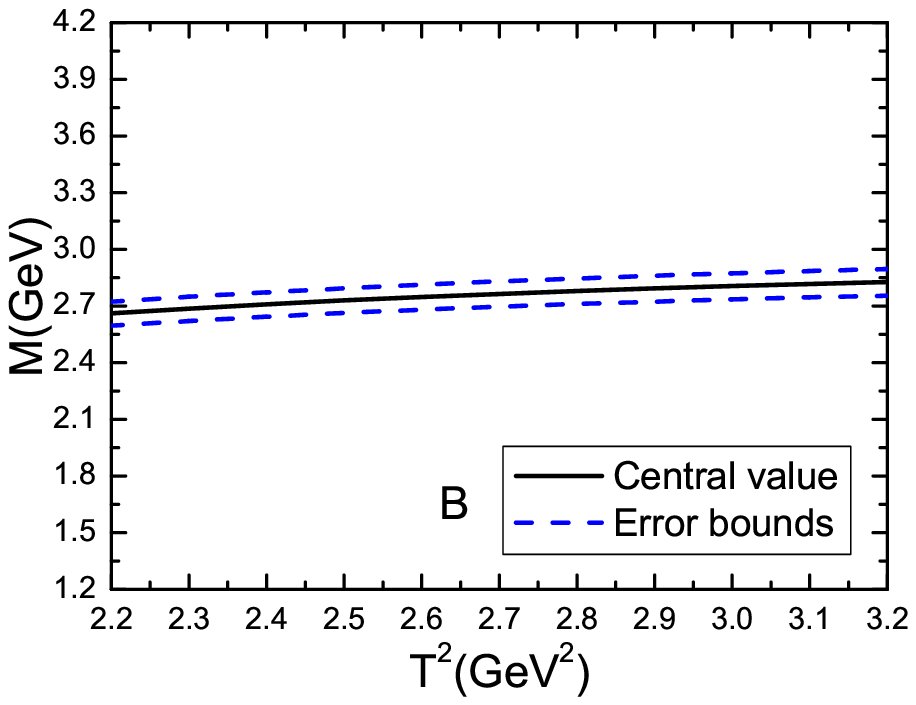}
 \includegraphics[totalheight=5cm,width=7cm]{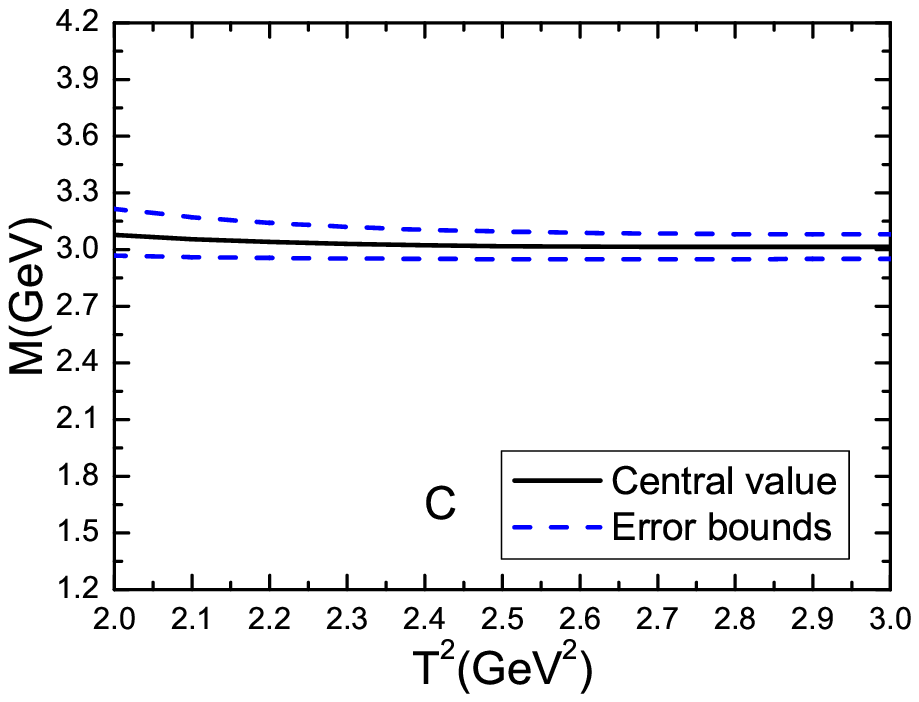}
 \includegraphics[totalheight=5cm,width=7cm]{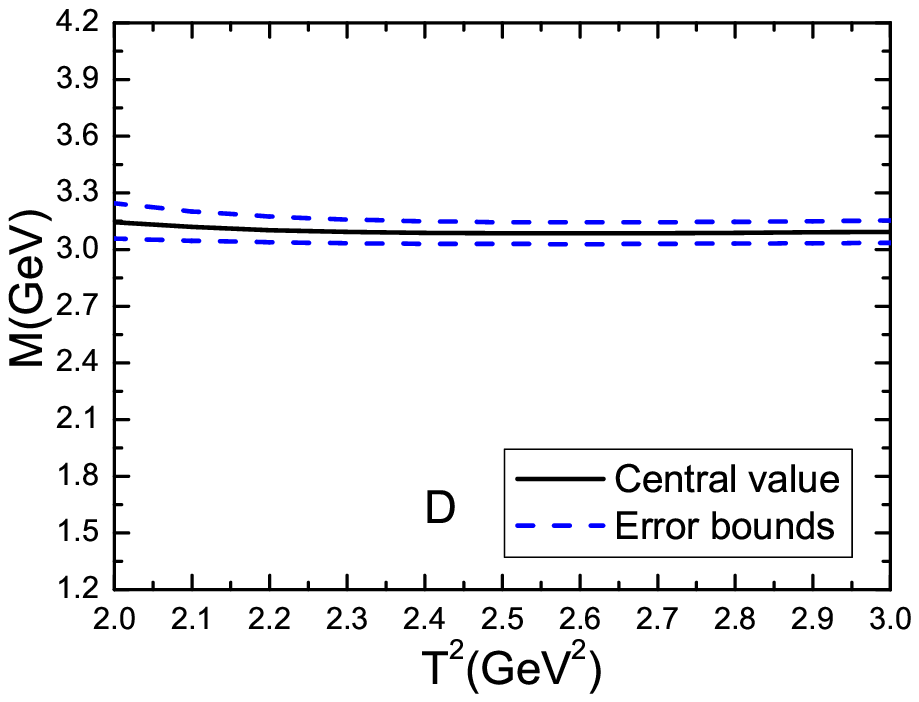}
 \includegraphics[totalheight=5cm,width=7cm]{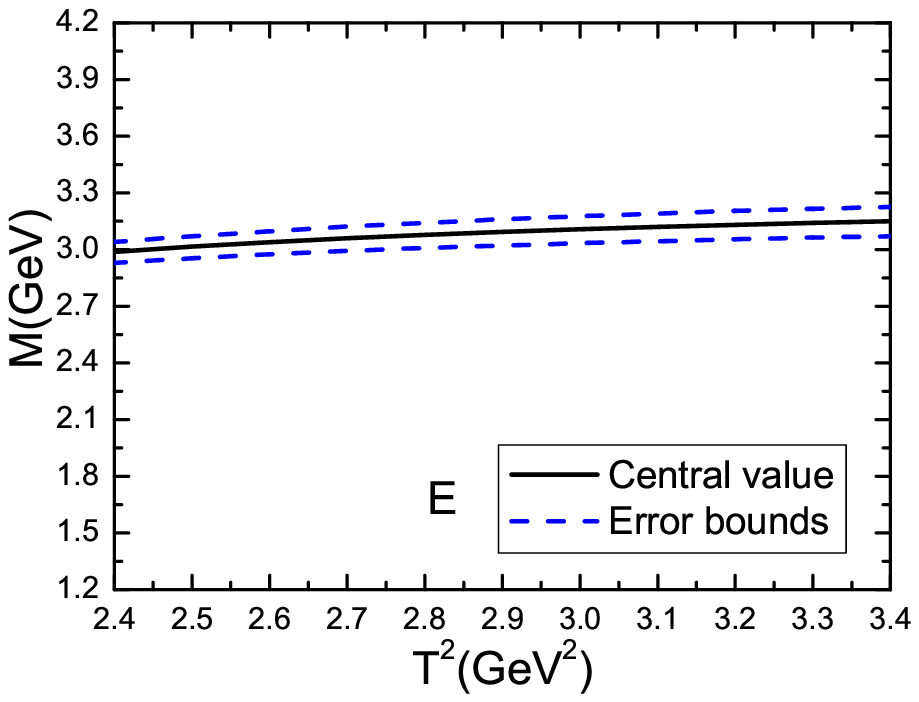}
 \includegraphics[totalheight=5cm,width=7cm]{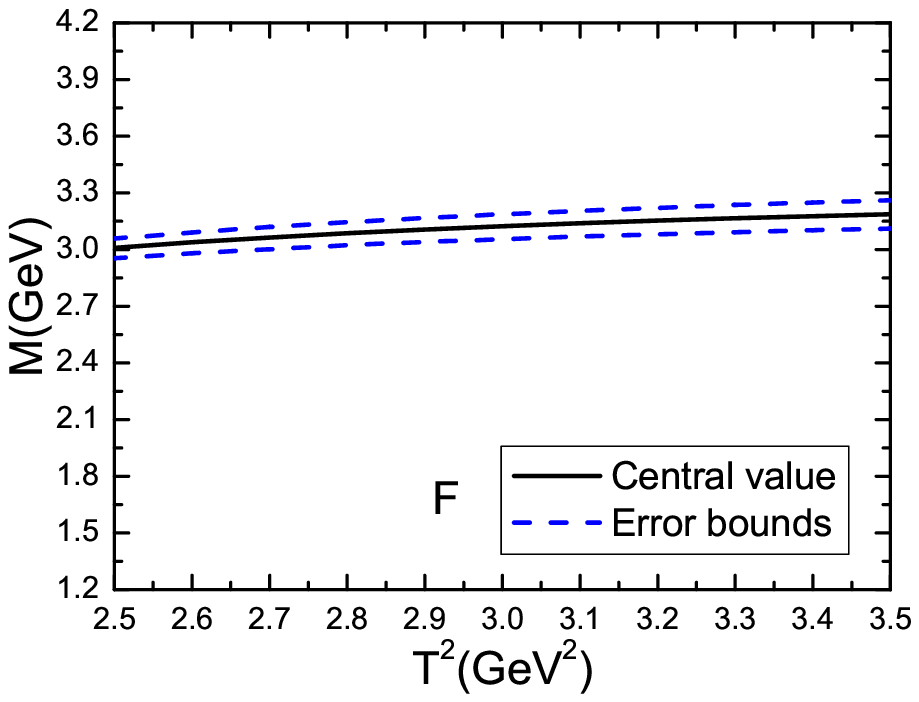}
 \includegraphics[totalheight=5cm,width=7cm]{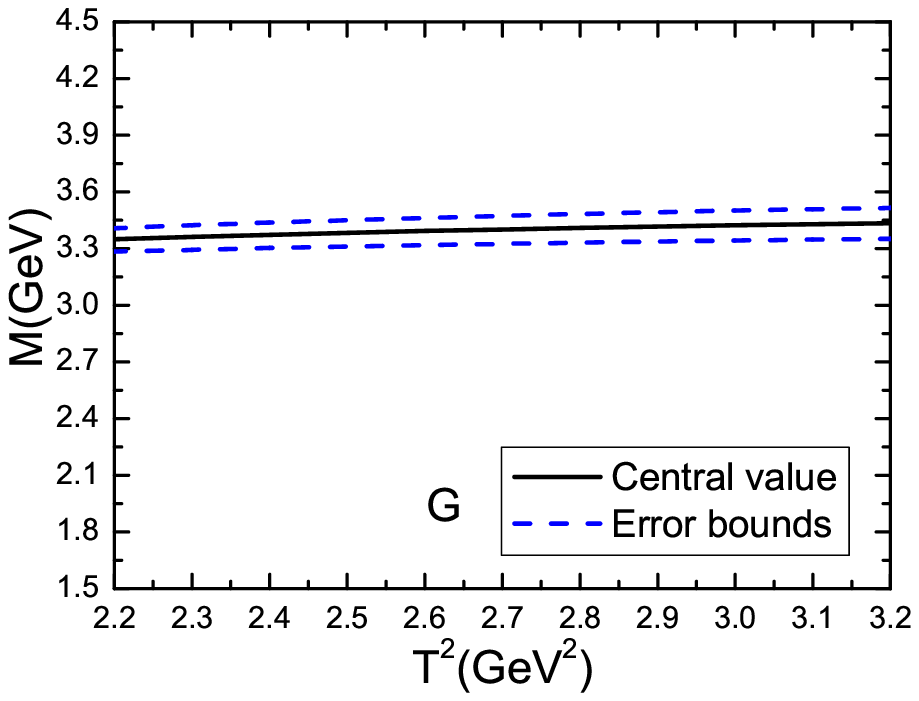}
 \includegraphics[totalheight=5cm,width=7cm]{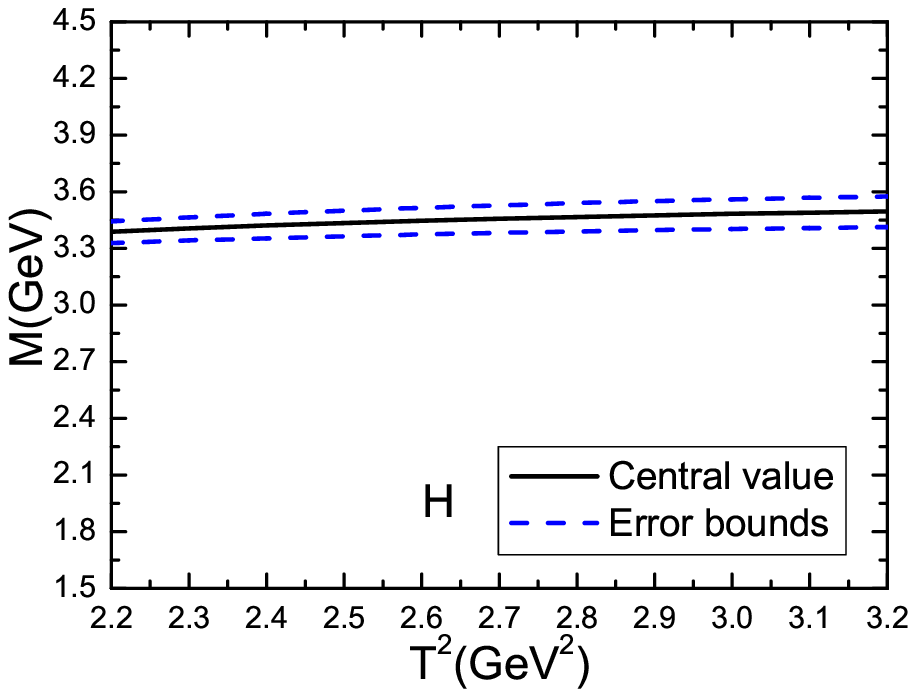}
         \caption{ The masses  of the $\Omega_c$ states   with variations of the Borel parameters $T^2$, where the $A$, $B$, $C$, $D$, $E$, $F$, $G$ and $H$
         correspond to the $\Omega_c$ states with the quantum numbers $\rm (1S,{\frac{1}{2}}^+)$, $\rm (1S,{\frac{3}{2}}^+)$,
         $\rm (1P,{\frac{1}{2}}^-)$, $\rm (1P,{\frac{3}{2}}^-)$,  $\rm (2S,{\frac{1}{2}}^+)$, $\rm (2S,{\frac{3}{2}}^+)$,
         $\rm (2P,{\frac{1}{2}}^-)$ and $\rm (2P,{\frac{3}{2}}^-)$, respectively.   }
\end{figure}

\begin{figure}
 \centering
 \includegraphics[totalheight=5cm,width=7cm]{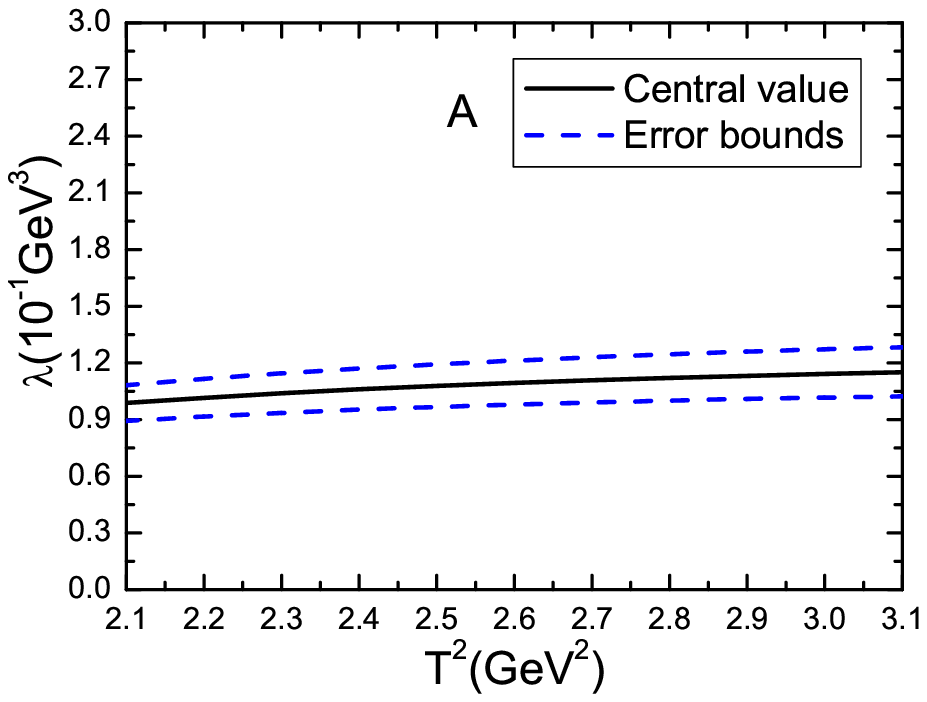}
 \includegraphics[totalheight=5cm,width=7cm]{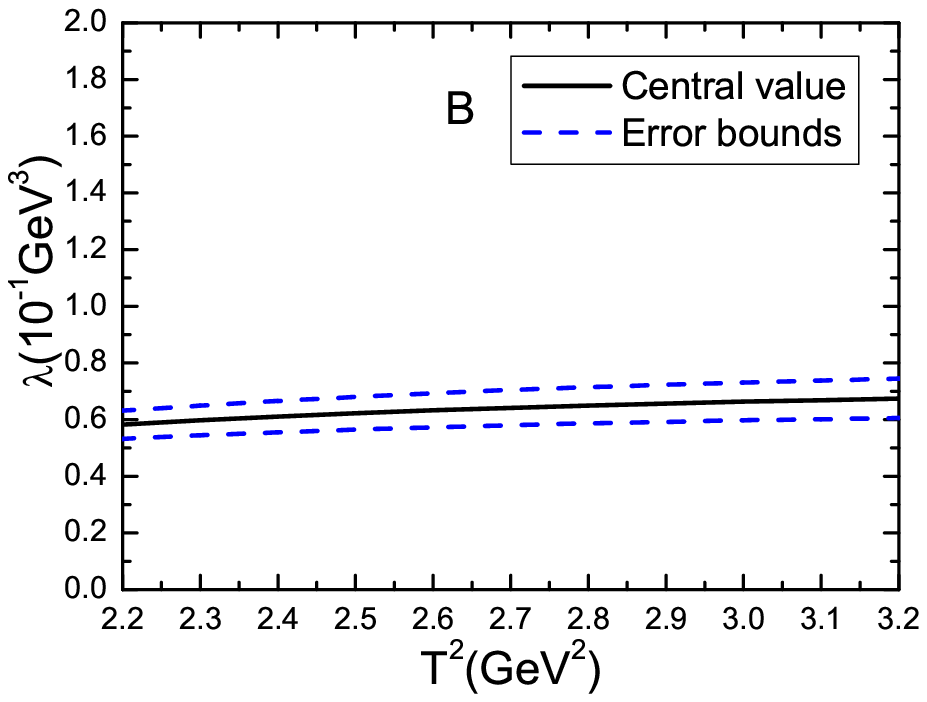}
 \includegraphics[totalheight=5cm,width=7cm]{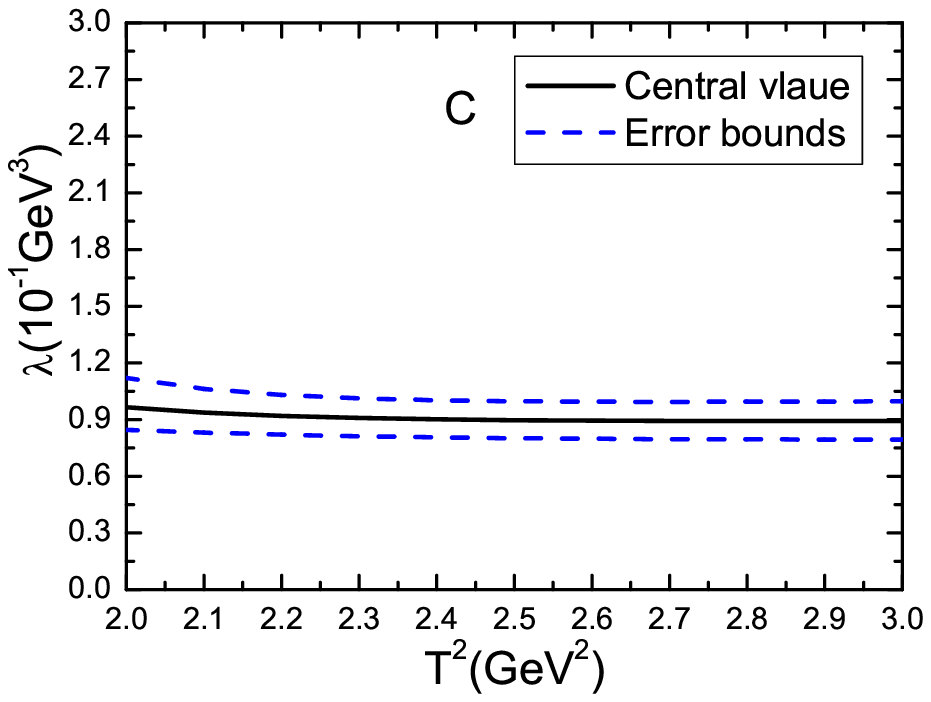}
 \includegraphics[totalheight=5cm,width=7cm]{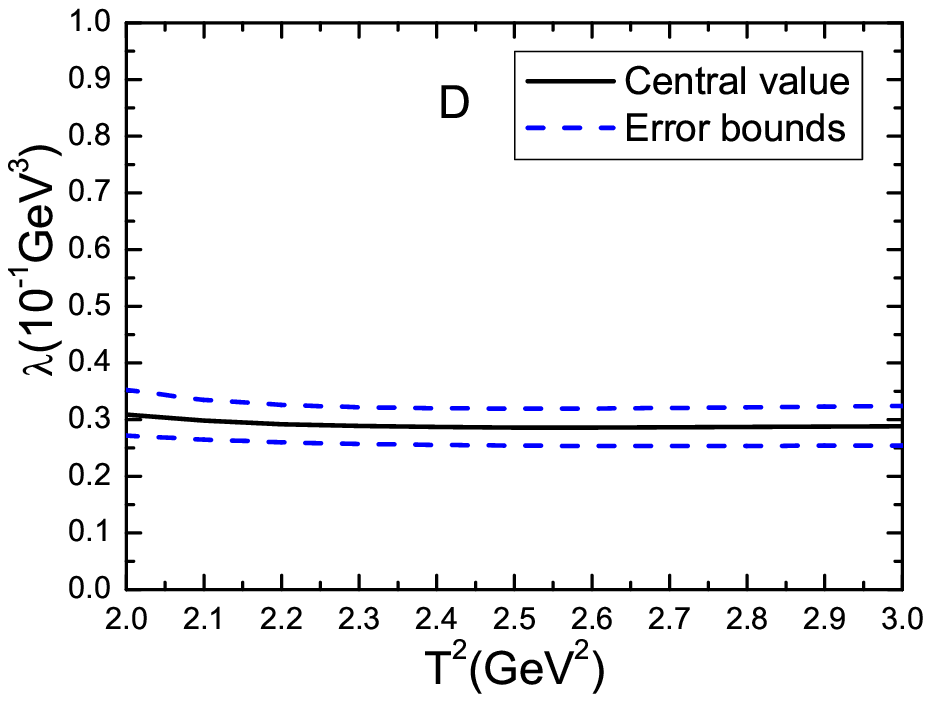}
 \includegraphics[totalheight=5cm,width=7cm]{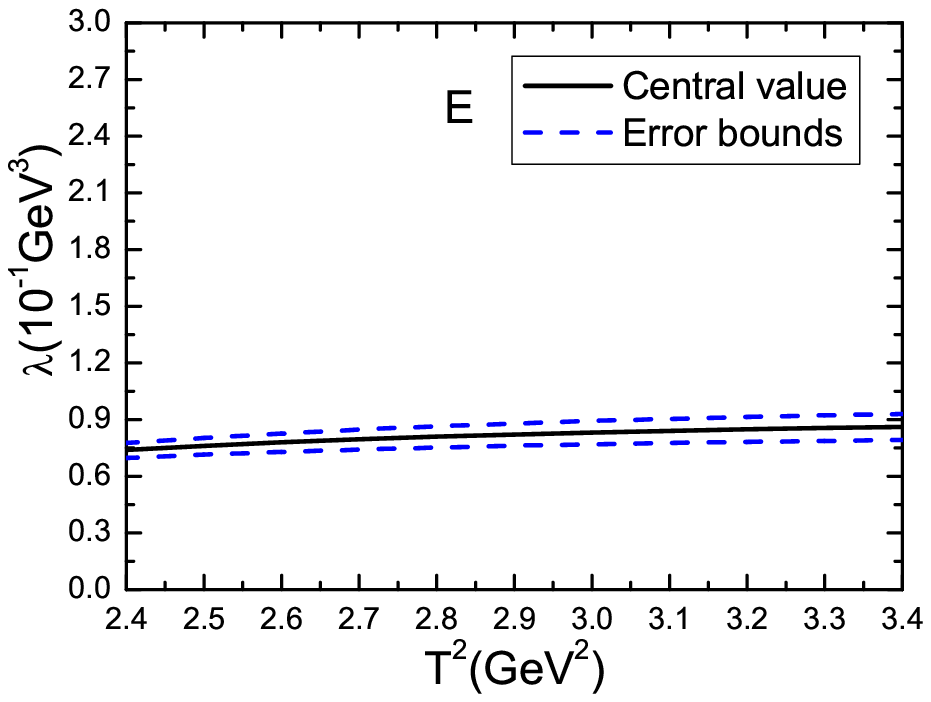}
 \includegraphics[totalheight=5cm,width=7cm]{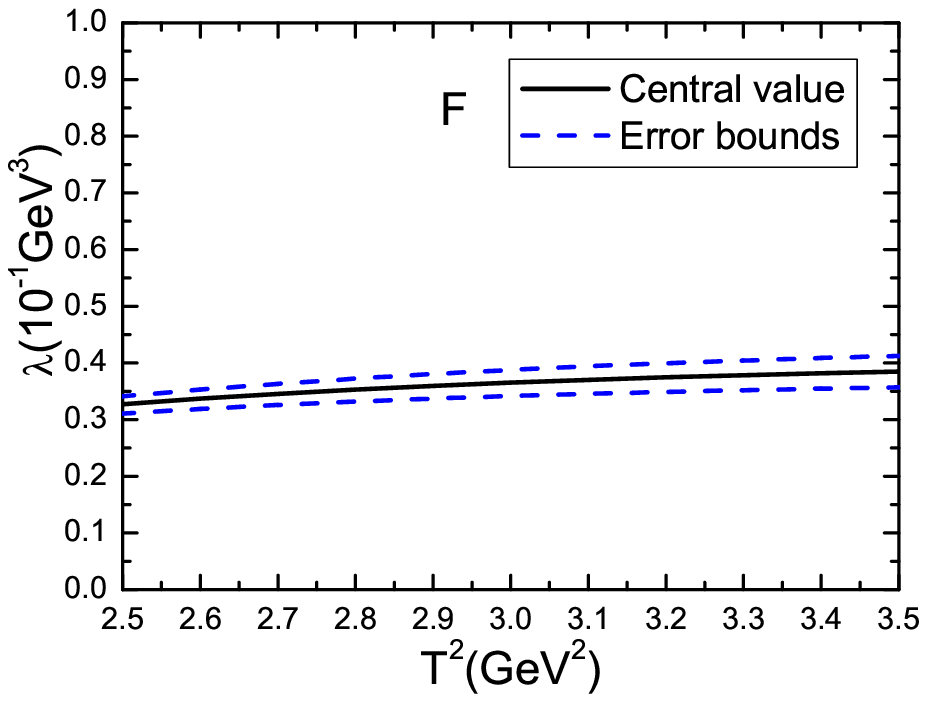}
 \includegraphics[totalheight=5cm,width=7cm]{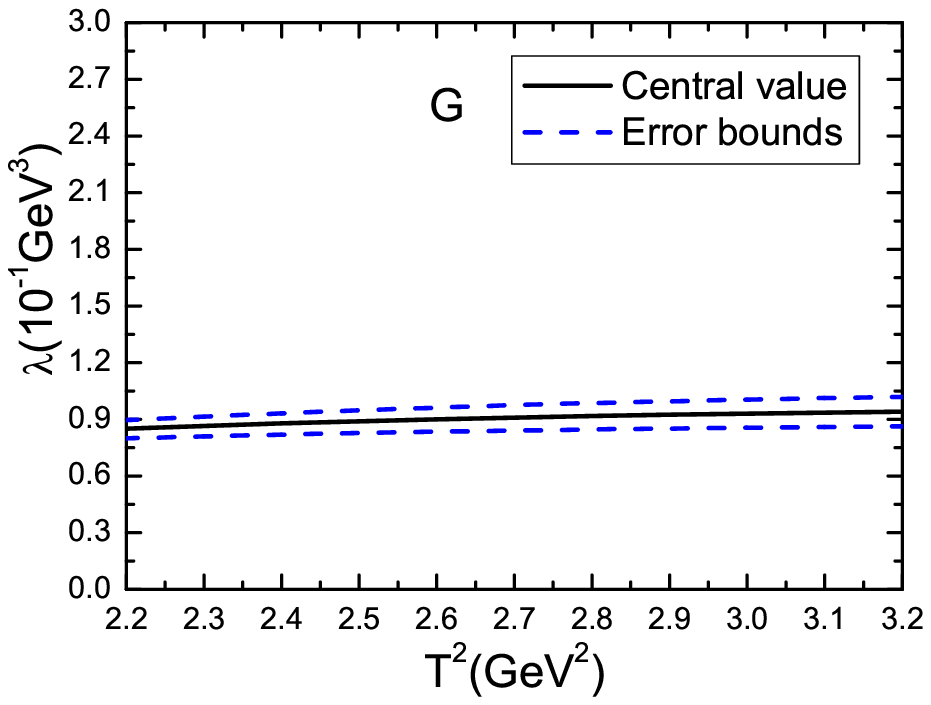}
 \includegraphics[totalheight=5cm,width=7cm]{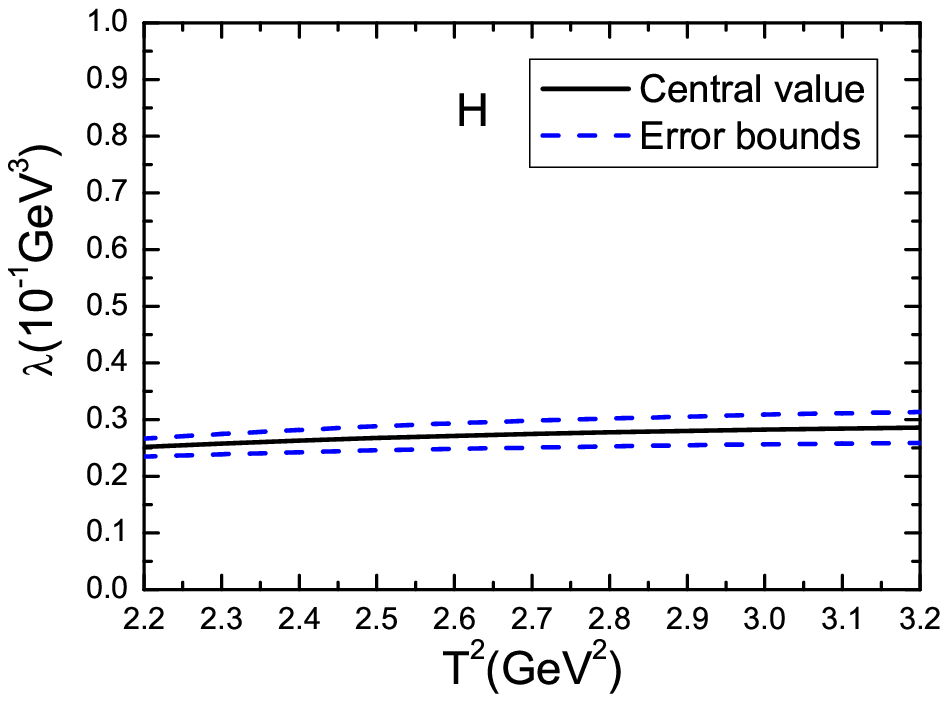}
         \caption{ The pole residues of the $\Omega_c$ states   with variations of the Borel parameters $T^2$, where the $A$, $B$, $C$, $D$, $E$, $F$, $G$ and $H$
         correspond to the $\Omega_c$ states with the quantum numbers $\rm (1S,{\frac{1}{2}}^+)$, $\rm (1S,{\frac{3}{2}}^+)$,
         $\rm (1P,{\frac{1}{2}}^-)$, $\rm (1P,{\frac{3}{2}}^-)$,  $\rm (2S,{\frac{1}{2}}^+)$, $\rm (2S,{\frac{3}{2}}^+)$,
         $\rm (2P,{\frac{1}{2}}^-)$ and $\rm (2P,{\frac{3}{2}}^-)$, respectively.  }
\end{figure}

\section{Conclusion}
In this article, we distinguish the contributions of the S-wave and P-wave $\Omega_c$ states unambiguously, study the masses and pole residues of the 1S, 1P, 2S and 2P
 $\Omega_c$ states with the spin $J=\frac{1}{2}$ and $\frac{3}{2}$ using the QCD sum rules in a consistent way, and revisit the assignments of the new narrow
 excited $\Omega_c$ states.
The present predictions support assigning the $\Omega_c(3000)$
 to be the 1P $\Omega_c$ state with $J^P={\frac{1}{2}}^-$, assigning the $\Omega_c(3090)$ to be the 1P $\Omega_c$ state with $J^P={\frac{3}{2}}^-$ or the
 2S $\Omega_c$ state with $J^P={\frac{1}{2}}^+$,   and assigning the $\Omega_c(3119)$  to be the 2S $\Omega_c$ state with $J^P={\frac{3}{2}}^+$.
 The present predictions indicate that the
1P $\Omega_c$ state with $J^P={\frac{3}{2}}^-$ and the 2S $\Omega_c$ state with $J^P={\frac{1}{2}}^+$ have degenerate masses,
it is difficult to distinguish them by the masses  alone, we have to study their  strong decays.  Other predictions   can be confronted to the experimental data in the future.

\section*{Acknowledgements}
This  work is supported by National Natural Science Foundation, Grant Number 11375063.

\end{document}